\documentclass[preprint,aps,preprintnumbers,amsmath,amssymb,nofootinbib,epsf,epsfig]{revtex4}
\usepackage[utf8]{inputenc}
\usepackage{graphicx}
\usepackage{amsmath}
\usepackage{amssymb}
\usepackage{array}
\usepackage{fancyhdr}
\usepackage{float}
\usepackage{booktabs}
\usepackage{multirow}
\usepackage{color}
\usepackage{caption}
\usepackage{subcaption}
\usepackage{natbib}
\usepackage[colorlinks, citecolor=cyan]{hyperref}
\usepackage{epsfig}
\usepackage{slashed}
\usepackage{appendix}
\usepackage{xcolor}
\usepackage{array}
\newcolumntype{C}[1]{>{\centering\arraybackslash}p{#1}}
\usepackage{slashed,dsfont}
\usepackage{bm,wasysym}
\usepackage{tabularx}
\usepackage{rotating}
\usepackage{lscape} 
\usepackage[normalem]{ulem}
\usepackage{placeins}
\newcommand{\lsim}{
\mathrel{\hbox{\rlap{\hbox{\lower4pt\hbox{$\sim$}}}\hbox{$<$}}}}

\newcommand{\gsim}{
\mathrel{\hbox{\rlap{\hbox{\lower4pt\hbox{$\sim$}}}\hbox{$>$}}}}

\allowdisplaybreaks[2]

\newcommand{\nn}{\nonumber}

\def\re{{\rm Re}}  \def\im{{\rm Im}}

\definecolor{schrift}{RGB}{120,0,0}

\def\apasl{{A_{\parallel \rm S}^L}}
\def\apasr{{A_{\parallel \rm S}^R}}  
\def\apaol{{A_{\parallel 0}^L}} 
\def\apaor{{A_{\parallel 0}^R}}
\def\apanl{{A_{\parallel 1}^L}} 
\def\apanr{{A_{\parallel 1}^R}}
\def\bpanl{{B_{\parallel 1}^L}}
\def\bpanr{{B_{\parallel 1}^R}}
\def\apatl{{A_{\parallel t}^L}}
\def\apatr{{A_{\parallel t}^R}}

\def\apesl{{A_{\perp \rm S}^L}}
\def\apesr{{A_{\perp \rm S}^R}}
\def\apeol{{A_{\perp 0}^L}}
\def\apeor{{A_{\perp 0}^R}}
\def\apenl{{A_{\perp 1}^L}} 
\def\apenr{{A_{\perp 1}^R}} 
\def\bpenl{{B_{\perp 1}^L}} 
\def\bpenr{{B_{\perp 1}^R}}
\def\apetl{{A_{\perp t}^L}}
\def\apetr{{A_{\perp t}^R}}
\usepackage{pifont}

\newcommand{\blue}[1]{{\color{blue} #1 }}

\newcommand{\be}{\begin{equation}}
\newcommand{\ee}{\end{equation}}
\newcommand{\bea}{\begin{eqnarray}}
\newcommand{\eea}{\end{eqnarray}}

\def\s1{\hat s}

\begin{document}
\title{\bf Correlative study of flavor anomalies and dark matter in the light of scalar leptoquark}

\author{Manas Kumar Mohapatra$^a$}
\email{manasmohapatra12@gmail.com}
\author{Shivaramakrishna Singirala$^{a,b}$}
\email{krishnas542@gmail.com}
\author{Dhiren Panda$^a$}
\email{pandadhiren530@gmail.com}
\author{Rukmani Mohanta$^a$}
\email{rmsp@uohyd.ernet.in}
\affiliation{$^a$School of Physics,  University of Hyderabad, Hyderabad-500046, India\\$^b$School of Physical Sciences, Indian Association for the Cultivation of Science, 2A \& 2B Raja S.C Mullick Road, Kolkata-700032, India}


\begin{abstract}
We explore $U(1)_{L_e-L_\mu}$ gauge extension of the Standard Model with particle content enlarged by three neutral fermions, of which the lightest one contributes to dark matter content of the Universe. The scalar sector is enriched with a $\tilde{R}_2$ scalar leptoquark doublet to investigate flavor anomalies in $B$-meson sector and a scalar singlet to spontaneously break the new $U(1)$. We discuss dark matter relic density and direct detection cross section in scalar and gauge portals. 
On the other hand, the new physics contribution for $b \to s$ transition comes from penguin diagrams with $Z^\prime$, leptoquark and new fermions. We analyze the constraints on the model parameters from the established observables of $B \to K^{(*)} \mu^+ \mu^-$ and $B_s\to \phi \mu^+ \mu^-$ decay channels. Utilizing the permissible parameter space consistent with both flavor and dark sectors, we discuss the impact on various observables such as branching ratio, forward-backward asymmetry and longitudinal polarisation asymmetry of $\Lambda_b \to \Lambda ^* (1520) (\to pK) \ell ^+\ell ^-$ decay channel.
\end{abstract}

\maketitle
\flushbottom
\section{Introduction} 
The Standard Model (SM) of  particle physics has been extremely triumphant in explaining the physics at elementary level. However, it fails to provide suitable explanation for several aspects.  Listing a few, the nature and existence of dark matter (DM) \cite{Zwicky:1937zza, Rubin:1970zza, Clowe:2003tk,Bertone:2004pz,ArkaniHamed:2008qn,Dodelson:1993je}, light neutrino masses and its oscillation phenomena \cite{Super-Kamiokande:1998kpq}, matter-antimatter asymmetry of the Universe \cite{Sakharov:1967dj,Kolb:1979qa, Davidson:2008bu,Buchmuller:2004nz,Strumia:2006qk} and anomalies associated with the $B$-meson sector.

Flavor-changing neutral current (FCNC) processes are important probes for physics beyond the SM. In particular, the $b \to s \mu^+ \mu^-$ transition has recently attracted considerable interest due to anomalies detected in LHCb~\cite{LHCb:2012bin, LHCb:2012juf, LHCb:2014cxe, LHCb:2014auh,LHCb:2014vgu,LHCb:2022vje} and Belle~\cite{Belle:2009zue, Belle:2019xld} experiments. The most compelling observables in $b \to s \ell \ell$ transitions, which have garnered significant attention for providing clear indications of new physics (NP), are the ratios $R_K$ and $R_{K^*}$ \cite{Hiller:2003js}, defined as
\begin{equation}
R_{K^{(*)}} = \frac{\mathrm{BR}(B \to K^{(*)} \mu \mu)}{\mathrm{BR}(B \to K^{(*)} e e)}\,,
\end{equation}
which are predicted approximately to be 1 in the SM, representing lepton flavor universality (LFU). They are theoretically clean as hadronic uncertainties cancel out to a large extent in these ratios.
Recent updates from LHCb \cite{LHCb:2022qnv, LHCb:2022vje} have confirmed that the measured values of these observables align with their SM predictions. Nevertheless, a range of other observables in $b \to s \ell \ell$ transitions, such as the notable $P_5'$ observable and various branching fractions, exhibit deviations of several sigma from SM predictions. Specifically, the LHCb \cite{LHCb:2013ghj,LHCb:2015svh} and ATLAS \cite{ATLAS:2018gqc} collaborations report a $3.3\sigma$ deviation in the measurement of $P_5'$ from the SM expectation. Additionally, the branching ratio for the $B_s \to \phi \mu^- \mu^+$ decay shows a $3.3\sigma$ deviation \cite{LHCb:2021zwz,LHCb:2015wdu} in the $q^2$ range of $[1.1, 6.0]$ $\rm GeV^2$. Furthermore, the measurements of $R_{K_S^0}$ and $R_{K^{*+}}$ \cite{LHCb:2021lvy} also exhibit deviations from their SM predictions of $1.4\sigma$ and $1.5\sigma$, respectively. Similarly,  various other observables such as forward-backward asymmetries, longitudinal polarization asymmetries, CP-averaged observables, CP asymmetries in \(B \to K^{(*)}\mu^+\mu^-\) decay processes across multiple $q^2$ intervals indicate few sigma deviations from their SM expectations. Consequently, the observations do not entirely exclude the possibility of new physics in FCNC-mediated $b \to s \ell \ell$ transitions. Various analyses, using both model-dependent and model-independent approaches, have been performed to account for the anomalies in $b \to s \ell^+ \ell^-$ transitions \cite{Singirala:2018mio,Singirala:2021gok,Rajeev:2021ntt, Rajeev:2020aut, Das:2023kch,Yadav:2024cbt,Mohapatra:2021izl,Dutta:2019wxo,Mohapatra:2021ynn,Sahoo:2016nvx,Bobeth:2001jm,Descotes-Genon:2020buf, Fajfer:2018bfj, Mohapatra:2024lmp,Yadav:2024rax,Bauer:2015knc,Das:2016vkr,Becirevic:2016yqi,Sahoo:2016pet,Hiller:2016kry,Crivellin:2015lwa,Ko:2017lzd,King:2017anf,DiChiara:2017cjq,Alonso:2017bff,Bonilla:2017lsq,Ellis:2017nrp,Ghosh:2022vpb, Greljo:2022jac, Guadagnoli:2023ddc, Altmannshofer:2023uci}.\\

To confirm these hints of NP in $b \to s\ell^+\ell^-$, it is crucial to exploit the growing dataset from the LHCb experiment, not only for more precise measurements of known observables but also to explore other decays probing the same underlying physics. In this regard, motivated by the observed discrepancies between the measured values and SM predictions in the above discussed transitions, we investigate the decay  of the $\Lambda_b$ baryon into excited $\Lambda$ state through the decay channel $\Lambda_b \to (\Lambda^* (1520) \to pK) \ell^+\ell^-$. This channel offers valuable insights into the dynamics of flavor transitions, shedding light on the underlying processes and enhancing our understanding of the implications of potential new physics. Among the semileptonic decay modes of $\Lambda_b$ to hadrons, the decay to the $\Lambda^*(1520)$ is particularly noteworthy due to its narrow width and its significance in pentaquark searches by the LHCb collaboration through the decay $\Lambda^0_b \to pK J/\psi$ \cite{LHCb:2015yax}. In addition, the $\Lambda^*$ characterized by its $J^P=3/2^-$ spin-parity and its strong decay into the $p\bar{K}$ pair, stands out from neighboring states such as the $\Lambda(1600)$, $\Lambda(1405)$, and the weakly decaying $\Lambda(1116)$, all of which have $J^P=1/2^\pm$. Moreover, this mode has attracted considerable interest due to its unique theoretical framework, despite lacking experimental observation. Measurements of the decay $\Lambda_b\to\Lambda(\to p \pi^-)\mu^+\mu^-$, including its differential branching fraction and angular observables, have been reported by several experiments \cite{CDF:2011buy,LHCb:2013uqx,LHCb:2015tgy,LHCb:2018jna}, along with the branching fraction of the radiative decay $\Lambda_b\to\Lambda\gamma$ \cite{LHCb:2019wwi}. A comprehensive study of $b\to s\mu^+\mu^-$ transitions, considering  all 33 independent angular observables of $\Lambda_b\to\Lambda(\to p \pi^-)\mu^+\mu^-$ have been analyzed, employing lattice QCD form factors for $\Lambda_b\to \Lambda$ \cite{Detmold:2016pkz}, and found to be consistent with both SM predictions and the anomalies seen in $B$ meson decays \cite{Blake:2019guk}.
A recent measurement from LHCb experiment  \cite{LHCb:2019efc} presents the first test of lepton universality in the baryon sector, achieved by measuring the ratio of branching fractions for the decays $\Lambda^0_b \to pK^-\mu^+\mu^-$ and $\Lambda^0_b \to pK^-e^+e^-$, referred to as $R_{pK}$.  While the ratio $R_{pK}$ is compatible with SM predictions, there is a noticeable suppression in $\mathrm{BR}(\Lambda_b \to pK \mu^+ \mu^-)$ relative to $\mathrm{BR}(\Lambda_b \to pK e^+ e^-)$. Therefore, interpreting this finding requires precise theoretical knowledge of the various excited $\Lambda$ states contributing to the broad $pK$ region, highlighting the necessity of developing a comprehensive understanding of these excited states. Taking this into account, it is quite intriguing to investigate the decay channel $\Lambda_b \to \Lambda^* (1520) (\to pK) \ell \ell$ within the Standard Model and beyond.

The invariant-mass distribution of the $pK^{-}$ system in the decay  $\Lambda_b \to pK^{-}\ell^{+}\ell^{-}$, for values of $q^{2}$ away from the $\phi$, $J/\psi$, and $\psi'$ resonances, is expected to be similar to the corresponding distribution observed with $q^2$ on-resonance. Such a distribution has already been studied in the process $\Lambda_b \to pK^{-}J/\psi(\to \ell^{+}\ell^{-})$~\cite{LHCb:2015yax}. As illustrated in Fig.~3 of Ref.~\cite{LHCb:2015yax}, a large number of excited 
$\Lambda^{*}$ baryon states can contribute within the overlapping mass 
regions. Among these, one state generates a distinct and narrow 
peak that is clearly separated from the others: the 
$\Lambda^{*}(1520)$ resonance. This state has a measured width of 
$15.6 \pm 1.0$ MeV~\cite{ParticleDataGroup:2024cfk} and is the lowest-mass resonance with quantum numbers $J^{P} = \frac{3}{2}^{-}$. Consequently, it maybe feasible for LHCb to measure the decay rate and angular observables of the process $\Lambda_b \to \Lambda^{*}(1520)(\to pK^{-})\ell^{+}\ell^{-}$ in the nonresonant region of $q^{2}$. Being specifically designed for precision studies of heavy flavor hadrons, the LHCb experiment provides excellent vertex resolution, efficient tracking, and powerful particle identification. This allows for a clean reconstruction of final states involving protons, kaons, and leptons. Therefore, the decay channel $\Lambda_b \to \Lambda^{*}(1520)(\to pK)\ell^{+}\ell^{-}$ particularly interesting to study within and beyond the SM.

The objective of this paper is to address flavor anomalies by realizing the new contributions to $b\to s$ transition via penguin diagrams with scalar leptoquark and a weakly interacting massive particle  (WIMP) dark matter flawing in the loop.
 It is well-known that the $U(1)_{L_\alpha - L_\beta}$, with ($\alpha, \beta= e,\mu,\tau$), gauge extension of the Standard Model is a very well motivated, simple and economical  BSM framework, which is  free from all the triangle anomalies. Motivated by the recent updated LHCb results on lepton flavor universality observables $R_{K^{(*)}}$ \cite{LHCb:2022qnv, LHCb:2022vje}, which are consistent with their corresponding Standard Model predictions, hinting towards the possibility of new contributions to both $b \to s \mu^+ \mu^-$ and $b \to s e^+ e^-$ channels,  we propose a new physics model based on a local $U(1)_{L_e-L_\mu}$ gauge extension of the SM to address the observed  anomalies associated with $b \to s \mu^+\mu^-$ transitions. By extending SM gauge symmetry with an additional $U(1)$, we obtain the new physics contribution in $B$-meson sector via $Z^\prime$ exchange, which are dictated by the  new gauge parameters and the Yukawa couplings. Moreover, the scalar leptoquark and $Z^\prime$ can also be the portal for dark matter (DM) annihilations to meet the relic density of the Universe and also provide spin-dependent and spin-independent WIMP-nucleon cross section. 
Among the scalar leptoquarks, we also consider the $\tilde{R}_2$ representation, which provides appropriate couplings to the quark and leptons and can contribute to both flavor observables and dark matter phenomenology. In particular, collider searches place constraints on the mass of $\tilde{R}_2$ depending on the strength of its Yukawa couplings. Updated LHC searches for the scalar $\tilde{R}_2$ leptoquarks \cite{CMS:2023qdw} exclude masses below 1.22 TeV for the Yukawa couplings considered here.  The $Z'$ boson parameter space is further constrained by ATLAS di-muon resonance searches \cite{ATLAS:2023vxg}.

This paper is structured as follows: Section II introduces the model, detailing the particle content, relevant interaction Lagrangian terms, scalar and fermion  mass matrices, and their diagonalization. Section III discusses the relic density of fermionic dark matter and the direct detection cross section prospects. 
In Section V and VI, we constrain the new physics parameter space from the flavor sector and explore the implications of new physics on the decay channel $\Lambda_b \to \Lambda^* (1520) (\to pK) \ell \ell$, respectively. Finally, our conclusions are summarized in Section VII.

\section{New $U(1)_{L_e-L_\mu}$ 
model with leptoquarks}
We consider a variant of $L_{e}-L_{\mu}$ model to address anomalies associated with $B$ meson decays and dark matter sector.  To obtain a suitable platform for a correlative study, we enrich the model with three neutral fermions $N_{e}, N_{\mu}, N_{\tau}$ and a scalar leptoquark $\tilde{R}_2$.  Alongside, a singlet $\phi_2$ is included to spontaneously break the new $U(1)$ gauge symmetry. On top, additional discrete $Z_2$ is imposed under which all the new fermions and $\tilde{R_2}$ are odd, helps to avoid unwanted vertices leading to dark matter decay and also allows suitable interactions for flavor anomalies. Similar works in the context of $L_{\mu}-L_\tau$ gauge extension can be looked at in \cite{Singirala:2018mio, Singirala:2021gok}. The full list of field content is provided in Table.  \ref{lelmu_model}.
\begin{table}[h!]
\begin{center}
\begin{tabular}{|c|c|c|c|c|}
	\hline
			& ~Field	& $ SU(3)_C \times SU(2)_L\times U(1)_Y$	~&~ $U(1)_{L_{e}-L_{\mu}}$	~& ~$Z_2$~\\
	\hline
	\hline
	Fermions		& $Q_L \equiv(u, d)^T_L$			& $(\textbf{3},\textbf{2}, 1/6)$	& $0$	& $+$\\
			& $u_R$							& $(\textbf{3},\textbf{1}, 2/3)$	& $0$ & $+$	\\
			& $d_R$							& $(\textbf{3},\textbf{1},-1/3)$	& $0$	& $+$\\
			& $\ell_{\alpha L} \equiv (\nu_\alpha, \alpha)_L$,	$\alpha = e,\mu,\tau$ & $(\textbf{1},\textbf{2},  -1/2)$	&  $1,-1,0$	& $+$\\
			& $\ell_R \equiv \alpha_R$, $\alpha = e,\mu,\tau$							& $(\textbf{1},\textbf{1},  -1)$	&  $1,-1,0$	& $+$\\
			& $N_{e},N_{\mu},N_{\tau}$						& $(\textbf{1},\textbf{1}, 0)$	&  $1,-1,0$	& $-$\\
	\hline
	Scalars	& $H$							& $(\textbf{1},\textbf{2},~ 1/2)$	&   $0$	& $+$\\
        
			& $\phi_2$						& $(\textbf{1},\textbf{1},~   0)$	&  $2$	& $+$\\  
			& $\tilde{R}_2$						& $(\textbf{3},\textbf{2},   1/6)$	&  $1$	& $-$\\    
	\hline
	\hline
\end{tabular}
\caption{Fields in the chosen $U(1)_{L_{e}-L_{\mu}}$ model.}
\label{lelmu_model}
\end{center}
\end{table}
The relevant Lagrangian terms corresponding to gauge, fermion, gauge-fermion interaction and scalar sectors  are given by
\begin{align}
{\cal L} &= {\cal L}_{\rm SM}  - \frac{1}{4} Z^{\prime}_{\mu\nu} Z^{\prime \mu\nu} - g_{e\mu} \overline{\ell_e}_L \gamma^\mu \ell_{eL} Z_\mu^\prime - g_{e\mu} \overline{e}_R \, \gamma^\mu e_R Z_\mu^\prime + g_{e\mu} \overline{\ell_\mu}_L \gamma^\mu \ell_{\mu L} Z_\mu^\prime + g_{e\mu} \overline{\mu}_R \, \gamma^\mu \mu_R Z_\mu^\prime \nn\\
    &+ \overline{N}_e \left( i \slashed{\partial} - g_{e\mu} Z_\mu^\prime \gamma^\mu \right) N_e + \overline{N}_\mu \left( i \slashed{\partial} + g_{e\mu} Z_\mu^\prime \gamma^\mu \right) N_\mu + \overline{N}_\tau i \slashed{\partial} N_\tau - \frac{f_e}{2} \left( \overline{N_e^c} N_e \phi_2^\dagger + \text{h.c.} \right) \nn\\
         &- \frac{f_\mu}{2} \left( \overline{N_\mu^c} N_\mu \phi_2 + \text{h.c.} \right) - \frac{1}{2} M_{\tau\tau} \overline{N_\tau^c} N_\tau - M_{e\mu} \left( \overline{N_e^c} N_\mu + \overline{N_\mu^c} N_e \right)  \nn\\
         &- \left( y_{qRN} \overline{Q}_L \tilde{R}_2 N_{\mu} + \text{h.c.} \right) + \left| \left( i \partial_\mu - \frac{g}{2} \boldsymbol{\tau}^a \cdot \mathbf{W}_\mu^a - \frac{g^\prime}{6} B_\mu + g_{e\mu} Z_\mu^\prime \right) \tilde{R}_2 \right|^2 \nn\\ &+ \left| \left( i \partial_\mu - 2 g_{e\mu} Z_\mu^\prime \right) \phi_2 \right|^2 
          - V\left (H, \tilde{R}_2,  \phi_2 \right ),
\label{eq:Lag}
\end{align}
where, the scalar potential is expressed as
\begin{align}
V(H,\tilde{R}_2,\phi_2) &=  \mu^2_{H}(H^{\dagger}H)+ \lambda_{H}(H^{\dagger}H)^2 +
  \mu^2_{\phi} (\phi^\dagger_2 \phi_2) + \lambda_{\phi} (\phi^\dagger_2 \phi_2)^2 
      +\mu^2_{R} ({\tilde{R}_2}^\dagger {\tilde{R}_2})  +\lambda_{R} (\tilde{R}_2^\dagger \tilde{R}_2)^2 \nonumber \\
      +&  \lambda_{H\phi} (\phi^\dagger_2 \phi_2)(H^\dagger H)+  \lambda_{R\phi}(\phi^\dagger_2 \phi_2) (\tilde{R}_2^\dagger \tilde{R}_2)  
      +  \lambda_{HR}(H^{\dagger}H)(\tilde{R}_2^{\dagger}\tilde{R}_2) + \lambda'_{HR}(H^{\dagger}\tilde{R}_2)(\tilde{R}_2^{\dagger}H).
\label{eq:potential}
\end{align}
In the above, the leptoquark doublet is denoted by   $\tilde{R_2} = \begin{pmatrix}
		 \tilde{R}_2^{2/3}		\\
		 \tilde{R}_2^{-1/3}	\\
	\end{pmatrix}$.
After spontaneous symmetry breaking the masses of the leptoquarks are denoted by	
\begin{eqnarray}
&&M_{R2/3}^2 = \mu_{R}^2  + \frac{ \lambda_{ HR}}{2} v^2 +  \frac{ \lambda_{R\phi}}{2} v^2_2\;, \nn\\
&&M_{R1/3}^2 = \mu_{R}^2  +  \left(\lambda_{HR} + \lambda'_{HR}\right)\frac{v^2}{2} +  \frac{ \lambda_{R\phi}}{2} v^2_2\;.\label{eq:4}
\end{eqnarray}
In the above Eq. (\ref{eq:4}),   $M_{R2/3}$ and $M_{R1/3}$ correspond to the masses of leptoquark components $\tilde{R}_2^{2/3}$, $\tilde{R}^{-1/3}_2$ respectively. The vacuum expectation value (vev) of the scalar $\phi_2$ is considered as $v_2$, while the vev of the SM Higgs is taken as $v$. We further assume  both the leptoquark components to be degenerate in mass, i.e.,  $M_{R2/3} = M_{R1/3} = M_{LQ}$ (for $\lambda^\prime_{HR} \simeq 0$), where $M_{LQ}$ is taken to be $1.2$ TeV in the whole analysis of the paper. Furthermore, the mass of associated gauge boson of new $U(1)$ symmetry is given as $M_{Z^\prime} = 2v_2 g_{e\mu}$. On the other hand, the loop level mixing of $Z-Z^\prime$ is only possible if the particles running in the loop couple to both $Z$ and $Z^\prime$
\cite{Gherghetta:2019coi}. From the particle content, only electron, muon and leptoquark can couple directly to gauge bosons and contribute to the mixing \cite{Gherghetta:2019coi}. 
We have evaluated the expected contribution, and found that the induced mixing parameter $\epsilon$ remains very small, typically of the order $\mathcal{O}(10^{-7})$ for $\mu' \sim m_\ell$ and $\mathcal{O}(10^{-4})$ for $\mu' \sim M_{LQ}$ where $\mu'$ and $m_\ell$ are the renormalization scale and mass of charged lepton, respectively. Given its negligible numerical impact on the observables considered, we have not included this effect in our analysis. However, this holds at one-loop order, and at higher loop-order the condition can be relaxed.
\subsection{Fermion and scalar spectrum}
The fermion and scalar mass matrices take the form
\begin{align}
	M_N
	=
	\begin{pmatrix}
		 \frac{1}{\sqrt{2}}f_{e}v_2	& M_{e\mu}	\\
		 M_{e\mu}	& \frac{1}{\sqrt{2}}f_{\mu}v_2				\\
	\end{pmatrix} \;,
    \quad M_S^2
	=
	\begin{pmatrix}
		 2 \lambda_H v^2   & {\lambda}_{H\phi} {v}v_2  \\
 {\lambda}_{H\phi} {v}v_2  & 2 \lambda_{\phi} v^2_2				\\
	\end{pmatrix} \;.
\end{align}
One can diagonalize the above mass matrices by $U_{\delta}^T M_{N} U_{\delta} = {\rm{diag}}~[M_{N1},M_{N2}]$, and $U_{\zeta}^T M_{S}^2 U_{\zeta} = {\rm{diag}}~[M_{H1}^2,M_{H2}^2]$, where \begin{align}
U_\theta
	=
	\begin{pmatrix}
		 \cos{\theta}	& \sin{\theta}	\\
		 -\sin{\theta}	& \cos{\theta}	\\
	\end{pmatrix},
	\label{massMatrix1}
\end{align}
with $\zeta = \frac{1}{2}\tan^{-1}\left(\displaystyle{\frac{\lambda_{H\phi} v v_2}{\lambda_\phi v^2_2 - \lambda_H v^2}}\right)$ and $\delta = \frac{1}{2}\tan^{-1}\left(\displaystyle{\frac{2M_{e\mu}}{(f_\mu - f_{e})(v_2/\sqrt{2})}}\right)$.\\\\
We denote the scalar mass eigenstates as $H_1$ and $H_2$, with $H_1$ is assumed to be observed Higgs at LHC with $M_{H_1} = 125.09$ GeV and $v= 246$ GeV. The mixing parameter $\zeta$ is taken minimal to stay with LHC limits on Higgs decay width. 
The signal strength of $H_1$ for a particular decay channel $H_1 \to X \overline{X}$, where $X$ is any SM particle such as photon, quark or lepton, is defined by
\bea
\mu_{X \overline{X}}= \frac{\sigma {\rm BR}(H_1 \to X \overline{ X})}{\sigma {[\rm BR}(H_1 \to X \overline{X})]_{\rm SM}}\;,
\eea
where $\sigma$ and ${\rm BR}(H_1 \to X \overline{ X})$ are the production cross section of $H_1$ and its branching fraction to $X \overline{X}$ decay channel. The denominator in the above equation represents the same quantity for  the SM Higgs particle. If the neutral boson $H_1$ is similar to the SM Higgs boson then according to LHC result the signal strength ratio $\mu_{X \overline{ X}}$ should be $>0.8$ \cite{ParticleDataGroup:2024cfk}. We have verified that, the value of $\zeta$ considered in our analysis consistent with the LHC limit. 

We indicate $N_1$ and $N_2$ to be the fermion mass eigenstates, with the lightest one ($N_1$) as the probable dark matter candidate in the present work. 
\section{Dark Matter phenomenology}
\subsection{Relic density}
\begin{figure}[thb]
\begin{center}
\includegraphics[width=0.3\linewidth]{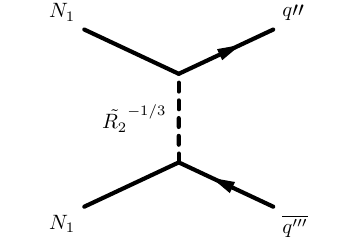}
\vspace{0.01 cm}
\includegraphics[width=0.3\linewidth]{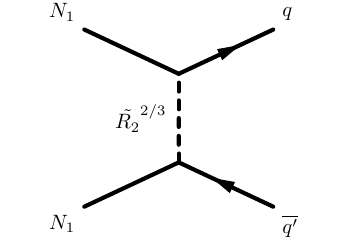}
\vspace{0.01 cm}
\includegraphics[width=0.3\linewidth]{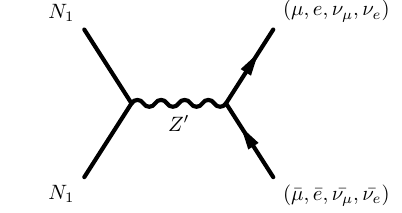}
\vspace{0.01 cm}
\includegraphics[width=0.3\linewidth]{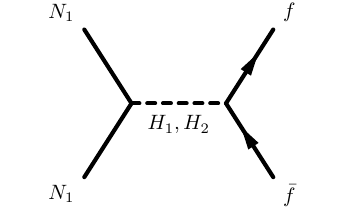}
\vspace{0.01 cm}
\includegraphics[width=0.3\linewidth]{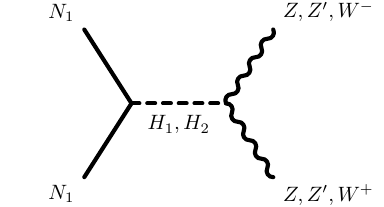}
\vspace{0.01 cm}
\includegraphics[width=0.3\linewidth]{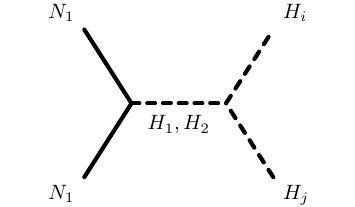}
    \caption{Feynman diagrams contributing to relic density.}
\label{relicfeyn}
\end{center}
\end{figure}

The model provides fermionic dark matter candidate $N_{1}$, with mass ranging up to $1$ TeV range. The DM fermion couples to LQ component ${\tilde R}_2^{-1/3}$ and can annihilate to $q^{\prime\prime} \bar{q^{\prime\prime\prime}}$ with $q^{\prime\prime}, q^{\prime\prime\prime} = d,s,b$. Also with ${\tilde R}_2^{2/3}$, it can annihilate to $q \bar{q^\prime}$ with $q, q^{\prime} = u,c,t$ in final state. In gauge portal, the DM can couple to $Z^\prime$ and can lead to s-channel annihilations with $e\overline{e}, ~\nu_e \overline{\nu_e},~ \mu \overline{\mu}, ~\nu_\mu \overline{\nu_\mu}$ as final state particles. Apart from these, CP-even portal i.e.,  $H_1,H_2$ mediated channels with $f\bar f, ZZ, Z^\prime Z^\prime, W^+W^-, H_i H_j$ in the final state can contribute to relic density,  where $f$ denotes all SM fermions and $i,j = 1,2$. The  corresponding Feynman diagrams are provided in Fig. \ref{relicfeyn}. The abundance of dark matter can be computed by \cite{PhysRevD.43.3191}
\begin{equation}
\label{eq:relicdensity}
\Omega h^2 = \frac{1.07 \times 10^{9} ~{\rm{GeV}}^{-1}}{  M_{\rm{Pl}}\; {g_\ast}^{1/2}}\frac{1}{J(x_f)}\;,
\end{equation}
where, $M_{\rm{Pl}}=1.22 \times 10^{19} ~\rm{GeV}$ and $g_\ast = 106.75$ denote the Planck mass and total number of effective relativistic degrees of freedom respectively. The function $J$ is \cite{PhysRevD.43.3191}
\begin{equation}
J(x_f)=\int_{x_f}^{\infty} \frac{ \langle \sigma v \rangle (x)}{x^2} dx.
\end{equation}
In the above, the thermally averaged cross section $\langle \sigma v \rangle$ reads  as \cite{Gondolo:1990dk}
\begin{equation}
 \langle\sigma v\rangle (x) = \frac{x}{8 M_{N1}^5 K_2^2(x)} \int_{4 M_{N1}^2}^\infty \hat{\sigma} \times ( s - 4 M_{N1}^2) \ \sqrt{s} \ K_1 \left(\frac{x \sqrt{s}}{M_{N1}}\right) ds.
\end{equation}
Here $K_1$, $K_2$ are the modified Bessel functions, $x = M_{N1}/T$, with $T$ being the temperature, $M_{N1}$ is dark matter mass, $\hat  \sigma$ is the dark matter cross section and $x_f$ stands for the freeze-out parameter. 
\begin{figure}[thb]
\begin{center}
\includegraphics[width=0.3\linewidth]{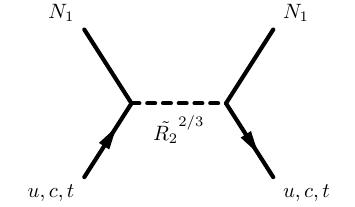}
\vspace{0.01 cm}
\includegraphics[width=0.3\linewidth]{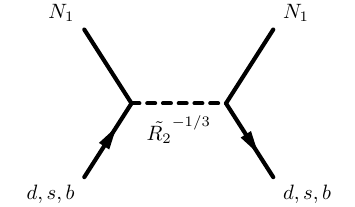}
\vspace{0.01 cm}
\includegraphics[width=0.3\linewidth]{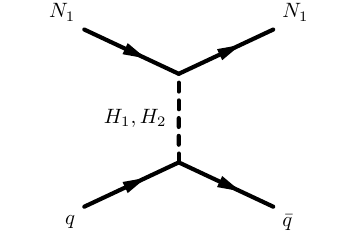}
\caption{Feynman diagrams contributing to WIMP-nucleon cross section.}
\label{DDfeyn}
\end{center}
\end{figure}
\subsection{Direct searches}
Moving to direct searches, the $Z^\prime$ does not directly couple to quarks, hence DM-nucleon cross section will not exist in gauge portal. However, the DM can couple to LQ components, will provide spin-dependent (SD) DM-neutron cross section, whose sensitivity can be checked with stringent upper bounds of LZ \cite{LZ:2022lsv} and XENON1T \cite{XENON:2019rxp}. The Feynman diagrams are provided in left and middle panels of Fig. \ref{DDfeyn} and the effective interaction Lagrangian takes the form
\begin{equation}
\mathcal{L_{\rm eff}} \simeq \frac{y_{qRN}^2\sin^2\delta}{4 M_{\rm LQ}^2} \overline{N_1}\gamma^\mu\gamma^5 N_1 ~\overline{q}\gamma_\mu\gamma^5 q\,.
\end{equation}
It should be noted that, in general  only the operators $\bar q \gamma_\mu q$ and $\bar q \gamma_\mu \gamma_5 q$  tend to dominate the cross section, and  of these the temporal component of
$\bar q \gamma_\mu q$ leads to spin-independent interactions, and the spatial component of $\bar q \gamma_\mu \gamma_5 q$ leads to spin-dependent interactions \citep{Agrawal:2010fh}. The spatial components of $\bar q \gamma_\mu  q$  and the temporal component of $\bar q \gamma_\mu \gamma_5 q$ are velocity suppressed. Therefore, the 
interactions of dark matter with nucleons for sizeable and spin-dependent interaction is due to the couplings of dark matter to quarks having the form $\bar q \gamma_\mu \gamma_5 q$. Additionally, for fermionic dark matter candidate one can have the interactions of the form $\bar N_1 \gamma^\mu N_1$ as well as  $\bar N_1 \gamma^\mu  \gamma_5 N_1$. However, for  Majorana fermion,  $\bar N_1 \gamma^\mu N_1$ term  identically vanishes and hence, only axial interaction survives. 

\begin{figure}[thb]
\begin{center}
\includegraphics[width=7.5cm,height=6cm]{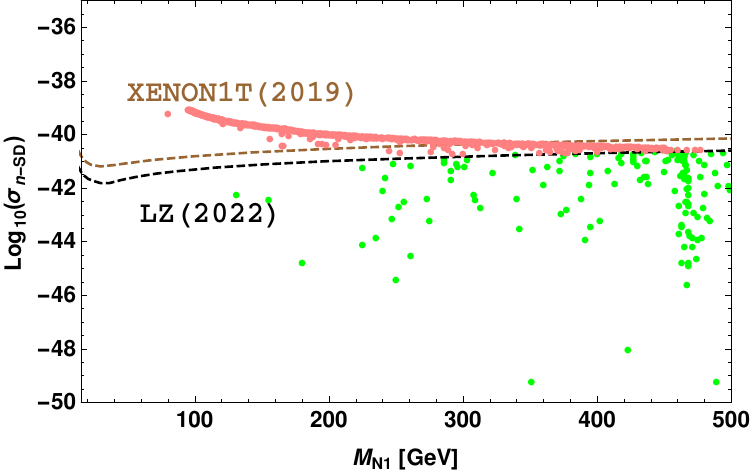}
\includegraphics[width=7.5cm,height=6.5cm]{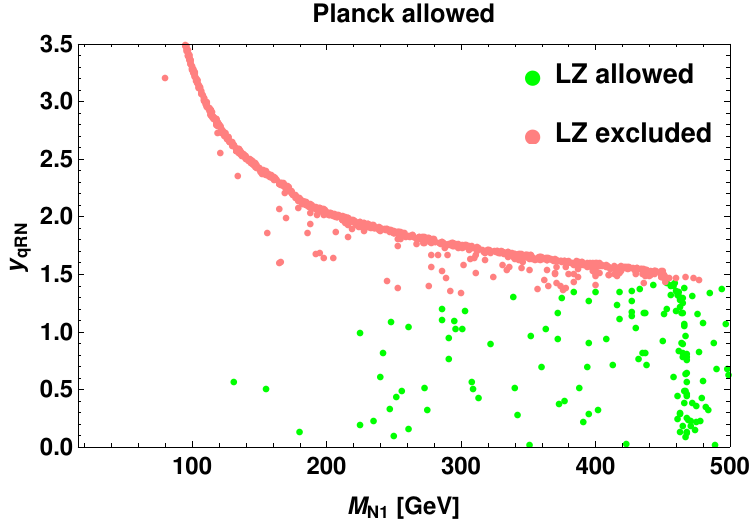}
\vspace*{0.2 true cm}
\includegraphics[width=7.5cm,height=6cm]{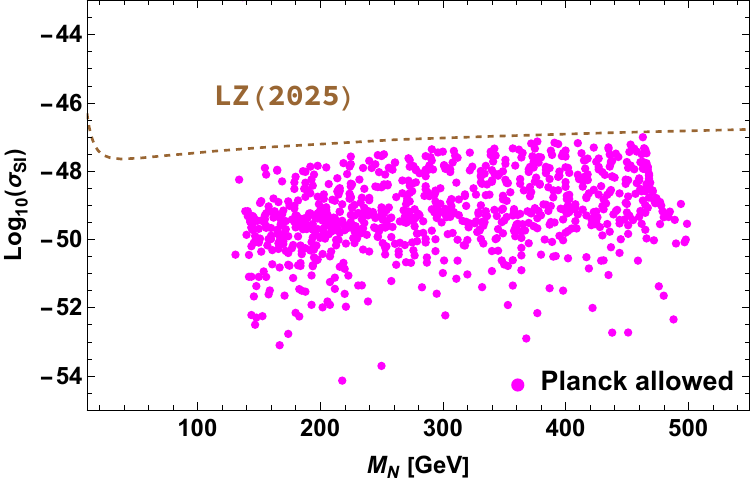}
\caption{WIMP-nucleon cross section as a function of dark matter mass projected in the top-left panel. Dashed lines correspond to upper limits of LZ \cite{LZ:2022lsv} and XENON1T \cite{XENON:2019rxp}. Top-right panel displays the favourable region in the plane of $y_{qRN}-M_{N1}$ plane showing same color contrast as the left panel. Bottom panel displays the spin-independent DM-nucleon cross section allowed by Planck  as a function of DM mass, where the dashed line represents the LZ limit \cite{LZ:2024zvo}.}
\label{DMDD}
\end{center}
\end{figure}

The spin-dependent cross section is given by \cite{Agrawal:2010fh}
\begin{equation}
\sigma_{\rm SD} = \frac{ \mu_r^2}{\pi} \frac{\sin^4\delta}{M_{\rm LQ}^4}y^4_{qRN}\left[\Delta_u + \Delta_d + \Delta_s\right]^2 J_n(J_n+1),
\end{equation}
where, $J_n = \frac{1}{2}$ is the angular momentum, reduced mass being $\mu_r = \frac{M_{N1} M_n}{M_{N1}+M_n}$ with $M_n \simeq 1$ GeV for nucleon. Values of quark spin functions $\Delta_{q}$ are given in \citep{Agrawal:2010fh}. Fig.~\ref{DMDD}, top-left panel, shows the allowed parameter space, along with the exclusion limits from \cite{LZ:2022lsv} and XENON1T \cite{XENON:2019rxp}.

In addition to the SD interaction, the spin-independent (SI) DM-nucleon cross section contributions can also arise in the Higgs portal, and are typically suppressed compared to the SD interaction. The effective Lagrangian
 in the Higgs portal can be read as
\begin{equation}
    \mathcal{L}_{\rm eff} \simeq a_q \overline{N_1^c}{N_1} \overline{q} q\;,  
\end{equation}
where, 
\begin{equation}
a_q = \frac{M_q}{v}(f_e \cos^2\delta + f_\mu \sin^2\delta) \left(\frac{\sqrt{2}\sin\zeta}{M_{H1}^2} - \frac{\sqrt{2}\cos\zeta}{M_{H2}^2} \right).    
\end{equation}

The Feynman diagram is shown in the right most panel of Fig. \ref{DDfeyn} and the spin-independent WIMP-neucleon cross section reads as
\begin{equation}
\sigma_{\rm SI} = \frac{4 \mu^2_r}{\pi}
f_p^2\;, 
 \label{sigmaSI}
\end{equation}

where, the hadronic matrix element $f_p$ is given as \cite{Okada:2012sg}
\begin{equation}
 \frac{f_p}{M_p} = \sum_{q=u,d,s}f_{Tq}^{p}\frac{a_q}{M_q} 
  + \frac{2}{27}\left(1-\sum_{q=u,d,s}f_{Tq}^{p}\right)\sum_{q=c,b,t}\frac{a_q}{M_q}\;.
\end{equation}
The values of $f_{Tq}$ for each quark flavor are provided in \cite{Ellis:2000ds}. We include the corresponding bounds from LZ limit~\cite{LZ:2024zvo} and the resulting constraints are taken into account in determining the allowed parameter space shown in bottom panel of Fig.~\ref{DMDD}.  We have implemented the model in LanHEP \cite{Semenov:1996es} package and used micrOMEGAs \cite{Pukhov:1999gg, Belanger:2006is, Belanger:2008sj} to compute relic density and also DM-nucleon cross section. 
\begin{figure}[thb]
\begin{center}
\includegraphics[width=7.5cm,height=6cm]{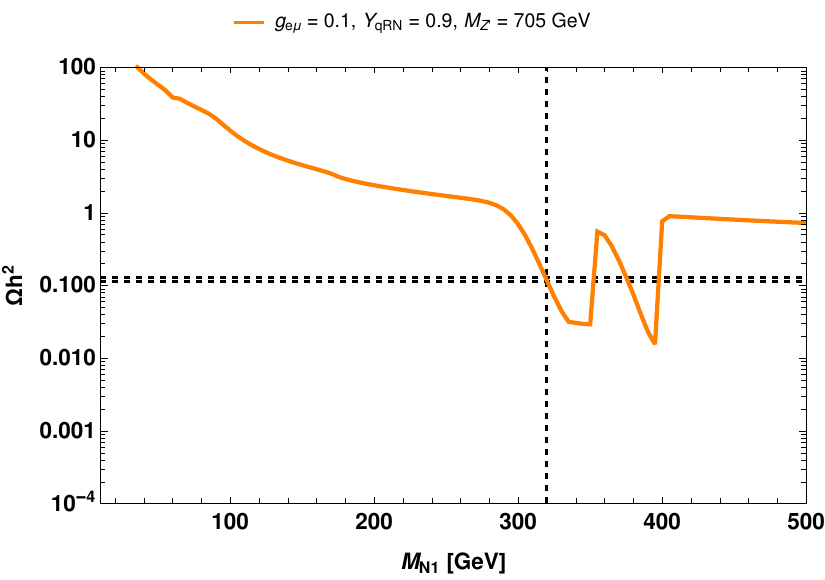}
\hspace{0.5 cm}
\includegraphics[width=7.5 cm,height=6 cm]{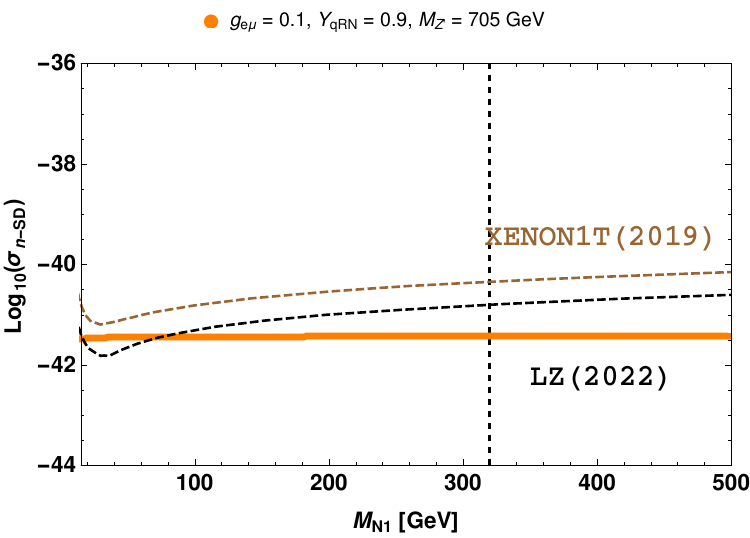}
\caption{Left panel projects relic density as a function of dark matter mass with horizontal dashed lines correspond to Planck $3\sigma$ constraint on relic density \cite{Aghanim:2018eyx}. Right panel depicts spin-dependent DM-neutron cross section with upper limits representing LZ \cite{LZ:2022lsv} and XENON1T \cite{XENON:2019rxp}. The input parameters are provided in Table.  \ref{benchmark}.}
\label{DMbench}
\end{center}
\end{figure}
\subsection{Analysis}
Top-left panel of Fig. \ref{DMDD} projects the SD DM-proton cross section as a function of DM mass. The red data points seem to violate the stringent LZ \cite{LZ:2022lsv} limit. The right panel shows that the allowed parameter space in $y_{qRN}-M_{N1}$ plane and indicates that Yukawa is indeed constrained by color contrast.  The bottom panel displays the spin-independent DM-nucleon cross section mediated through Higgs portal, allowed by Planck data as a function of DM mass, where the dashed line represents the LZ limit \cite{LZ:2024zvo}. We can also notice from these plots that the low-mass range (below 100 GeV) of DM is not allowed.

Left panel of Fig. \ref{DMbench} projects DM relic density meeting Planck satellite data \cite{Aghanim:2018eyx} for a $320$ GeV dark matter and other parameter values are provided in Table. \ref{benchmark}. First dip is due to s-channel resonance in $Z^\prime$ propagator with mass $M_{Z^\prime} = 705$ GeV. Second dip is due to resonance in $H_2$ propagator i.e., $M_{H2} = 800$ GeV. A small dip around $60$ GeV is because of resonance in $H_1$ propagator. Right panel displays SD DM-neutron cross section for the same $320$ GeV dark matter and of the benchmark provided in Table. \ref{benchmark}.

\section{Constraints on new physics from the flavor sector}

The most general effective Hamiltonian mediating the $ b \to s \ell^+ \ell^-$  transition is given by \cite{Bobeth:1999mk, Bobeth:2001jm} 
\bea
{\cal H}_{\rm eff} &=& - \frac{ 4 G_F}{\sqrt 2} V_{tb} V_{ts}^* \Bigg[\sum_{i=1}^6 C_i(\mu) O_i +\sum_{i=9,10} \Big ( C_i(\mu) O_i
+ C_i^\prime(\mu) O_i^\prime \Big )
\Bigg]\;,\label{ham}
\eea
where $G_F$ is the Fermi constant,  $V_{tb}V_{ts}^*$ denote the CKM matrix elements, $C_i$'s stand for  the Wilson coefficients evaluated at the renormalized scale $\mu = m_b$ \cite{Buras:1994dj}.

 Here $O_i$'s represent dimension-six operators responsible for leptonic/semileptonic processes, given as
\bea
O_9^{(\prime)}&=& \frac{\alpha_{\rm em}}{4 \pi} (\bar s \gamma^\mu P_{L(R)} b)(\bar \ell \gamma_\mu \ell)\;,~~~~~~~ O_{10}^{(\prime)}= \frac{\alpha_{\rm em}}{4 \pi} (\bar s \gamma^\mu 
P_{L(R)} b)(\bar \ell \gamma_\mu \gamma_5 \ell)\;,
\eea
where, $\alpha_{\rm em}$ is  the fine-structure constant, $P_{L,R} = (1\mp \gamma_5)/2$ are the chiral projectors.

\begin{figure}[thb]
\begin{center}
\includegraphics[width=0.32\linewidth]{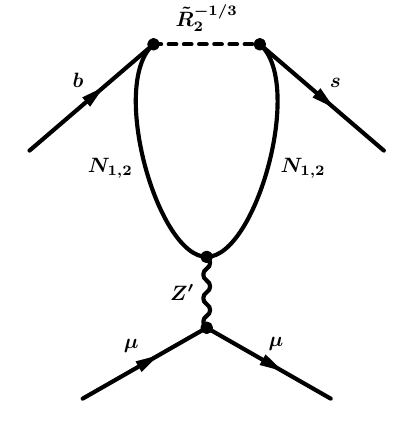}
\vspace{0.01 cm}
\includegraphics[width=0.32\linewidth]{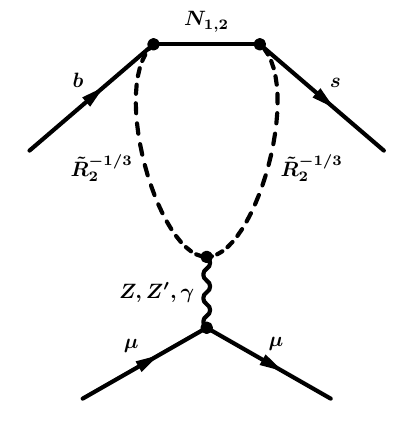}
\hspace{0.5 cm}
\includegraphics[width=0.32\linewidth]{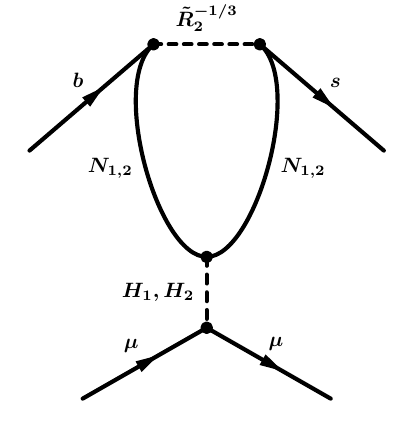}
\vspace{0.01 cm}
\includegraphics[width=0.32\linewidth]{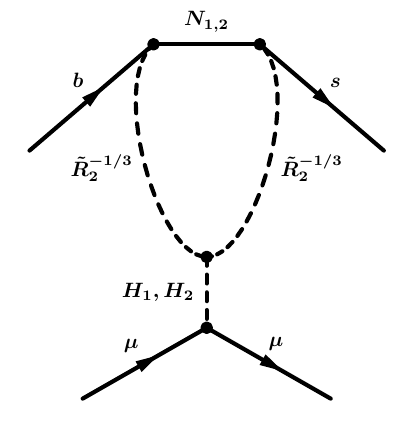}
\caption{Allowed penguin diagrams illustrating the $ b \to s \mu^+\mu^-$ transition in the model.}
\label{penguin}
\end{center}
\end{figure}

The one-loop diagrams contributing non-zero values to the rare $b \to s \ell \ell$ processes can arise through the exchange of $Z^\prime, H_{1,2}$  particles, forming penguin diagrams with the scalar leptoquark $\Tilde{R}_2^{-1/3}$ and $N_{1,2}$ particles within the loop, as depicted in Fig.  \ref{penguin}\,.  In the present framework, the one-loop penguin diagram illustrated in the top-left panel significantly contributes to the process. However, the diagram corresponding to the top-right panel yields  negligible contribution to the $b \to s \ell \ell$ transition as it is suppressed by $m_b/M_{N1}$ and $m_b/M_{N2}$. The loop functions of the lower two diagrams are suppressed by the factor $m_qM_{N{1}}/M_{LQ}^2$ and $m_qM_{N{2}}/M_{LQ}^2$ ($q=b, s$)  resulting in minimal contribution to $b \to s\ell \ell$ processes. Thus, only the top-left diagram, mediated via $Z^\prime$ boson, substantially contributes to the $b \to s \ell\ell$ channels.
The diagrams with propagators $N_{1}$ and $N_{2}$ exhibit superficial logarithmic divergences. These divergences are managed by employing a renormalization procedure followed by $\overline{\text{MS}}$ scheme. The renormalization scale $\mu'$ is chosen to be near the characteristic energy scale of the process under consideration. Specifically, we set $\mu'$ close to $M_{\text{LQ}}$ to minimize the magnitude of the logarithmic terms. We observe that the dependence on $\mu'$ is mild, indicating that the logarithmic term has a negligible impact on our results.

In the presence of $Z^\prime$ exchanging one loop diagram, the transition amplitude of semileptonic $b \to s \ell^+ \ell^-$ decay process  is given as 
\bea \label{amp}
\mathcal{M}=-\frac{1}{(4 \pi)^2}\frac{y_{q R N}^2 g_{e \mu}^2 }{4} \frac{\mathcal{R} (a, b)}{M_{Z^{\prime}}^2}[\bar{s}(p^{\prime})\gamma^\ell P_Lb(p)][\bar{\ell}(q_2)\gamma_\ell \ell(q_1))],\label{loop}
\eea
which in comparison  with the generalized effective Hamiltonian provides additional  new  Wilson coefficient as
\bea
C_9^{ \rm NP } &=& -\frac{1}{4 \pi} \frac{\sqrt{2}}{4G_F M_{Z^\prime}^2} \frac{1}{\alpha _{em}} \frac{y_{q R N}^2 g_{e \mu}^2 }{V_{tb}V_{ts}^*} \mathcal{R} (a, b),
\eea
where the loop function is given as
\bea
\mathcal{R} (a, b)&=\Bigg[ \frac{3}{2} \Big\{  \Big( \frac{1}{2} - \frac{\sqrt{a}\sqrt{b}}{a-b}\bigg( \frac{a \log a}{a-1} -\frac{b \log b}{b-1}  \Big) - \frac{1}{2 (a-b)} \big( \frac{a^2 \log a}{a-1} -\frac{b^2 \log b}{b-1}  \Big) \Big\} \nn\\
&- \frac{3}{2}\Big( \frac{1}{8} -\frac{3-3b^2+8b \log b -2b^2 \log b}{8(b-1)^2} \Big) +\frac{1}{2}\Big( \frac{1}{8} -\frac{3-3a^2+8a \log a -2a^2 \log a}{8(a-1)^2} \Big) \Bigg]\;,
\eea
with $a= \frac{M_{N1}^2}{M_{LQ}^2}$ and $b= \frac{M_{N2}^2}{M_{LQ}^2}$. Here, the mixing angle between the $(N_1, N_2)$ fermions, which is taken to be $30^\circ$ as a representative choice to illustrate the numerical results. Additionally, the mass splitting between $N_1$ and $N_2$ is considered to be 200 GeV.

Focusing on the study of NP couplings in the muon sector, our objective is to constrain the model parameters associated with leptoquark (LQ) and $Z^\prime$ couplings by analyzing decay modes involving $b \to s \mu^+ \mu^-$ transitions, specifically the $B \to K \mu^+ \mu^-$, $B \to K^* \mu^+ \mu^-$, and $B_s \to \phi \mu^+ \mu^-$ processes. We consider the associated decay observables, and carry out our analysis utilizing the \emph{flavio} package~\cite{Straub:2018kue}. 
We include the branching ratios for $B^0 \to K^{*0}\mu^+\mu^-$, $B^+ \to K^{*+}\mu^+\mu^-$, $B^0 \to K^0\mu^+\mu^-$, and $B^+ \to K^+\mu^+\mu^-$~\cite{LHCb:2014cxe, LHCb:2016ykl}, across various $q^2$ bins. We also include measurements of the forward-backward asymmetry ($A_{FB}$)~\cite{LHCb:2015svh}, longitudinal polarization fraction ($F_L$), CP asymmetries ($A_{i=3, 4, 5, 7, 8, 9}$), and form factor-independent observables ($P_1-P_8^{\prime}$)~\cite{LHCb:2020lmf, LHCb:2020gog} across multiple $q^2$ intervals such as $[0.10, 0.98]$, $[1.1, 2.5]$, $[2.5, 4.0]$, $[4.0, 6.0]$, and $[1.1, 6.0]$. 
Additionally, the updated branching fraction for $B_s \to \phi \mu^+ \mu^-$ as measured by LHCb over different $q^2$ intervals~\cite{LHCb:2021zwz} are used for this analysis. On the other hand, the latest measured values of the observables $R_K$ and $R_{K^*}$ are in agreement with the SM result \cite{LHCb:2022qnv}. However, we have included their contributions which are also relevant in the new physics scenario under consideration.
In our  analysis of constraining the parameter space, we  further considered the $B_s - \bar{B}_s$ mixing, which has  been presented in the following subsection.
\subsection{$B_s-\bar B_s$ mixing}
In this subsection, we examine the constraints imposed on the new parameters by the mass difference between the $B_s$ meson mass eigenstates ($\Delta M_s$), which serves as a key indicator of $B_s - \bar{B}_s$ mixing. Within the SM, this mixing occurs predominantly via a box diagram involving an internal top quark and $W$ boson exchange. The effective Hamiltonian governing the $\Delta B = 2$ transition is expressed as follows \cite{Inami:1980fz} 
\bea  \label{ham2}
{\cal H}_{\rm eff}=\frac{G_F^2}{16 \pi^2}~ \lambda_t^2~ M_W^2 S_0(x_t)\eta_B
(\bar s b)_{V-A}(\bar s b)_{V-A}\;, 
\eea
 where  $\lambda_t=V_{tb} V_{ts}^*$ and $\eta_B$ are the CKM matrix elements and the QCD correction factor, respectively.  The loop function $S_0(x_t)$ is given by  \cite{Inami:1980fz} 
 \be S_0(x_t)=\frac{4 x_t -11 x_t^2
+x_t^3}{4(1-x_t)^2} - \frac{3}{2} \frac{\log x_t x_t^3}{(1-x_t)^3}\;,
\ee 
 with $x_t=m_t^2/M_W^2$. 
Using Eqn. (\ref{ham2}),  the $B_s - \bar B_s$ mass difference in the SM is given as 
 \bea
\Delta M_s^{\rm SM} = 2 |M_{12}^{\rm SM}|  =\frac{\langle \bar B_s|{\cal H}_{eff}| B_s \rangle}{
M_{B_s}}  = \frac{G_F^2}{6
\pi^2} M_W^2~ \lambda_t^2~ \eta_B~ \hat B_s f_{B_s}^2 M_{B_s} S_0(x_t)\;.
\label{sm}
 \eea 
Using the given input parameters including the decay constant of the $B_s$ meson, $f_{B_s}$, the most recent SM prediction for $\Delta M_s$, reads from \cite{Albrecht:2024oyn}, is:
\bea \label{SM-Bs}
\Delta M_s^{\rm SM} = (18.23\pm 0.63)~ {\rm ps^{-1}},
\eea
and the most recent experimentally measured value reported in \cite{HeavyFlavorAveragingGroupHFLAV:2024ctg}, is given by
 \bea
\Delta M_s^{\rm Expt} = 17.765(6)~ {\rm ps^{-1}}.\label{Exp-Bs}
\eea
Although the theoretical prediction aligns well with the experimental data on $B_s - \bar{B}_s$ oscillations, the existence of new physics cannot be entirely ruled out. 
 \begin{figure}[thb]
 \begin{center} 
\includegraphics[width=0.42\linewidth]{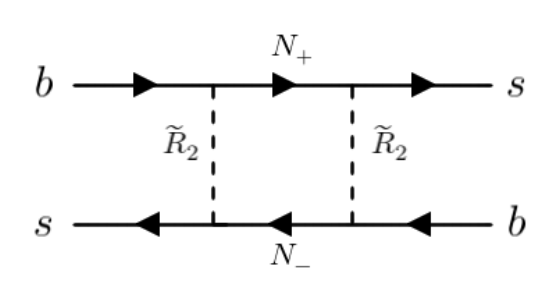}
\includegraphics[width=0.42\linewidth]{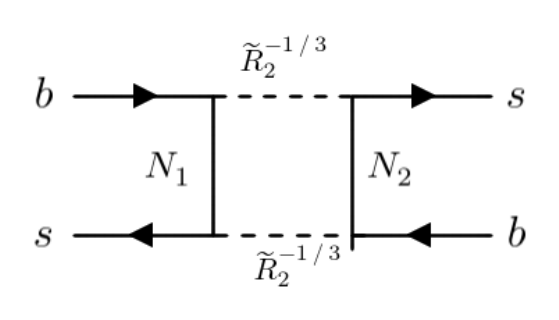}
\includegraphics[width=0.42\linewidth]{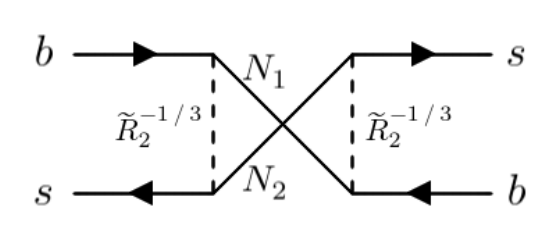}
\includegraphics[width=0.42\linewidth]{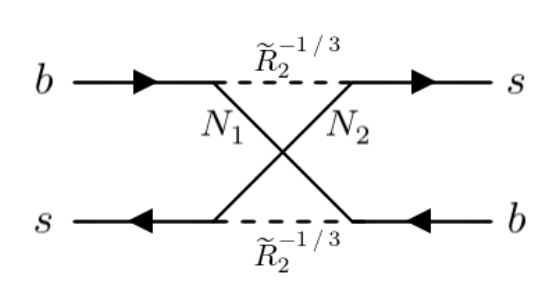}
 \caption{Box diagrams of $B_s-\bar B_s$ mixing  with leptoquark in the loop.}
 \label{box-Bs}
 \end{center}
 \end{figure}
The $B_s-\bar B_s$ mixing phenomena receives additional contributions in this model  through one-loop box diagrams involving the leptoquark $\tilde{R}_2$ and $N_{1,2}$ flowing inside the loop and are given in Fig.~\ref{box-Bs}. The effective Hamiltonian in the presence NP is given by \cite{Singirala:2018mio, Baek:2018aru, Baek:2017sew}
\bea \label{ham3}
{\cal H}_{\rm eff}=\frac{(y_{s RN} y_{b RN})^2}{128\pi^2 M_{LQ}^2} \cos^2 \delta \sin^2 \delta  C_{B_s}^{\rm NP}\left(\bar s b\right)_{V+A}\left(\bar s b\right)_{V+A}\;,
\eea 
where $\delta$ denotes mixing angle between the $(N_1, N_2)$ fermions. The loop function is given by  \cite{Singirala:2018mio,Baek:2018aru, Baek:2017sew}
\bea
C_{B_s}^{\rm NP}&=&2k\left(\chi_1, \chi_1,1\right)+ 4k\left(\chi_1, \chi_2,1\right)+2k\left(\chi_2, \chi_2,1\right)+\chi_1 j\left(\chi_1, \chi_1,1\right)\nn \\&+&2\sqrt{\chi_1\chi_2}j\left(\chi_1, \chi_2,1\right)+\chi_2j\left(\chi_2, \chi_1,1\right),
\eea
with $\chi_{1,2}=M_{N1, N2}^2/M_{LQ}^2$ and the loop functions $k\left(\chi_{1,2}, \chi_{1,2},1\right),~j\left(\chi_{2,1}, \chi_{1,2},1\right)$ are given in Appendix \ref{loop}.  
Now using Eqn. (\ref{ham3}), the mass difference of $B_s-\bar B_s$ mixing due to the exchange of the scalar doublet $\tilde{R}_2$ and the $N_{1,2}$ is found to be
\bea
\Delta M_s^{\rm NP}  = \frac{(y_{s RN} y_{b RN})^2}{48  \pi^2 M_{LQ}^2} \cos^2 \delta \sin^2 \delta \hspace*{0.05 true cm} C_{B_s}^{\rm NP} \eta_B \hat B_{B_s} f_{B_s}^2 M_{B_s}\;.
\label{LQ}
 \eea 
Including the NP contribution arising due to the scalar $\tilde{R}_2$ exchange, the  total mass difference can be written as
 \bea \label{Bs-delM-f}
\Delta M_s = \Delta M_s^{\rm SM}  \left [1 + \frac{C_{B_s}^{\rm NP}\cos^2 \delta \sin^2 \delta }{8 G_F^2 V_{tb}^2 V_{ts}^{*2} M_W^2 S_0(x_t)}
 \left (\frac{(y_{s RN} y_{b RN})^2}{M_{LQ}^2}  \right )\right ]\;.
\eea 
Using Eqns. (\ref{SM-Bs}) and (\ref{Exp-Bs}) in (\ref{Bs-delM-f}), one can put bound on  $y_{qRN}$ and $M_{N1}$ parameters.
The allowed parameter space in the $M_{Z^{\prime}} - g_{e\mu}$ and $M_{N1} - y_{qRN}$ planes has been depicted in the left and right panels of Fig.~\ref{constraint-flavor}, respectively, and is consistent with both dark matter and flavor constraints. In this analysis, we consider the benchmark values of the parameters as shown in Table \ref{benchmark}.
The intent of this figure is to illustrate the interplay between flavor constraints and dark matter considerations, highlighting how these two sets of constraints shape the allowed parameter space. This effectively distinguishes the regions defined by constraints from flavor and dark matter analyses, as well as their intersection while employing a refined color scheme for the clarity. Although the overall allowed region remains largely unchanged, the additional restrictions play a crucial role in refining the viable parameter ranges, highlighting the complementarity between these constraints. In this analysis, we consider that the coupling $y_{qRN}$ remains within the perturbative regime, specifically $|y_{qRN}| \lesssim \sqrt{4\pi}$. As observed in the left panel, the grey area is excluded by the experimental measurement of the neutrino trident production \cite{CCFR:1991lpl,Altmannshofer:2014pba}. Direct searches of $Z'$ bosons have been carried out at LEP-II as well. The coupling of the $Z'$ boson to leptons is tightly constrained by the measured 
cross section of the process $e^+e^- \to \ell^+\ell^-$ at the LEP-II experiment~\cite{Electroweak:2003ram}. For $M_{Z'} > \sqrt{s} = 209~\text{GeV}$, the limits can be expressed through four-fermion contact interaction bounds, leading to $g' \leq 0.044\, M_{Z'}/(200~\text{GeV})$~\cite{Buckley:2011vc}. 
In the region $M_{Z'} < 209~\text{GeV}$, the contact-interaction description is no longer valid, and a more conservative bound of $g' \leq 0.04$ is applied~\cite{Bell:2014tta}. These constraints are relevant to the relevant $U(1)_{L_e-L_\mu}$  scenario, illustrated by the light magenta shaded areas in Fig.~\ref{constraint-flavor}. For details, see Ref.~\cite{Dasgupta:2023zrh}. 

The common region in Fig. \ref{constraint-flavor} arises from various experimental constraints, such as bounds from neutrino trident production, LEP-II, rare decay processes and dark matter relic abundance. To further elucidate the role of these constraints, we emphasize that collider bounds play a crucial role in shaping the viable parameter space. In particular, LEP-II data impose strong limits in the low-mass region, while neutrino trident production provides additional constraints on the gauge coupling. These effects are already incorporated in Fig.~\ref{constraint-flavor}, where the allowed regions are consistent with all current experimental limits. In right panel, the parameter space allowed by $b \to s \mu^+\mu^-$ flavor observables and the dark matter relic abundance is shown purple points. However, the $B_s - \bar{B}_s$ mixing  imposes a stringent bound on the Yukawa coupling $y_{qRN}$, disfavoring the consistent region of $b \to s \mu^+\mu^-$ and dark matter. Consequently, the parameter space does not allow simultaneous explanation of all constraints. Additionally the decay observables of $\Lambda_b \to \Lambda^*(1520)(\to pK^-)\,\ell^+\ell^-$, due to the presence of stringent Yukawa coupling, are not significantly affected compared to the SM contributions. Therefore, we have included only the $b \to s \mu^+\mu^-$ observables that are consistent with dark matter relic abundance in this analysis. 
Our analysis includes a detailed description of how each of these constraints contributes to shaping the parameter space. Since we use experimental limits to exclude regions of parameter space rather than performing a statistical fit, so the confidence levels are not considered in this analysis.
\begin{figure}[thb]
\begin{center}
\includegraphics[width=8cm,height=6.2cm]{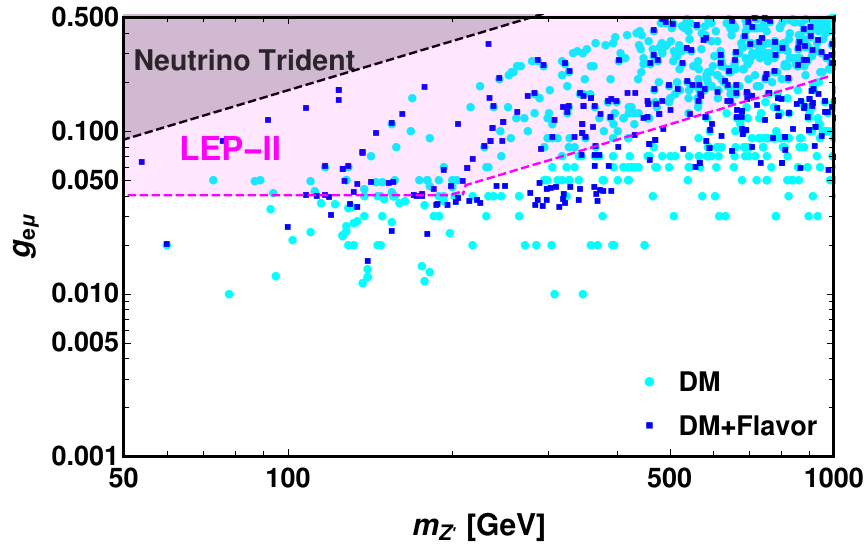}
\vspace{0.01 cm}
\includegraphics[width=8cm,height=6.5cm]{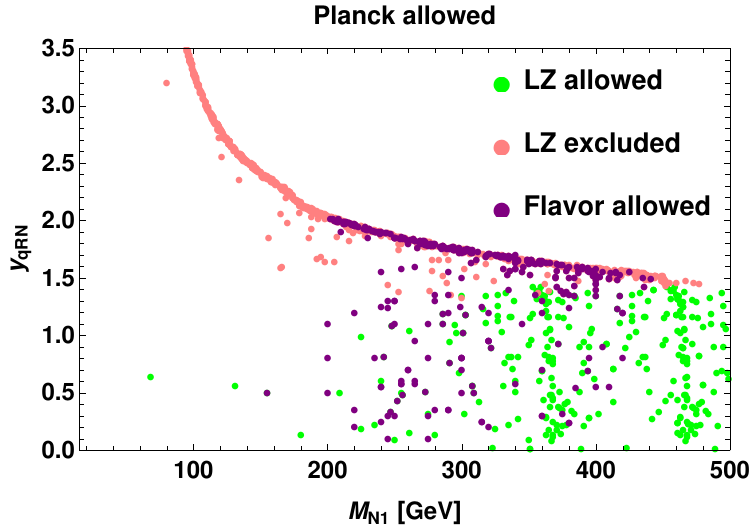}
\caption{Allowed parameter space from the $b \to s $ transitions and the dark sector. Left panel displays it in $g_{e\mu}$–$M_{Z^{\prime}}$ plane, incorporating the excluded regions from neutrino trident production \cite{CCFR:1991lpl,Altmannshofer:2014pba} and LEP-II bound \cite{Electroweak:2003ram,Dasgupta:2023zrh}, respectively. Right panel shows the constrained region in $y_{qRN}$–$M_{N1}$ plane. To mention, the leptoquark mass is taken to be $1.2$ TeV for the entire analysis.
}
\label{constraint-flavor}
\end{center}
\end{figure}
\begin{table}[htbp]
\caption{Values of the input parameters~\cite{ParticleDataGroup:2024cfk} used in our analysis.}
\centering
\scalebox{0.9}{
\begin{tabular}{|cc|cc|cc|cc|cc|}
\hline
Parameter & Value & Parameter & Value & Parameter & Value   \\
\hline
\hline
$G_F$ & $1.166 \times 10^{-5}$ GeV$^{-2}$ &$m_u$ & 0.003 GeV & $m_{\Lambda _b}$ & 5.619  GeV  \\
$\tau _{\Lambda _b}$ & $(1.470 \pm 0.010)\times 10^{-12}$ $s$  & $m_ c$ & $1.270$ GeV &$m_ {\Lambda ^*}$ & $1.519$ GeV\\
$|V_{tb}V_{ts}^*|$ & $0.0401 \pm 0.001$ & $m_b$ & 4.18 GeV & $m_\mu$ & 0.105  GeV\\
$\alpha _e$ &  1/127.925 & $\mathcal{B}_{\Lambda^*}$ & $0.45  \pm 0.01$  & $m_e$ & 0.0005 GeV \\
\hline
\end{tabular}}
\label{tab_input}
\end{table}
\begin{table}[htb]
\centering
\begin{tabular}{|c|c|c|c|c|c|c|c|}
\hline
\rm{Parameter} ~&~ ~$y_{qRN}$ ~&~$g_{e \mu}$~&~ $M_{Z^\prime}$ [GeV] ~&~ $M_{N1}$ [GeV]\\
\hline
\hline
\rm{Bechmark} ~&~ ~$0.9$~&~$0.1$~ &~$705$~&~$320$\\
\hline
\end{tabular}
\caption{Sample benchmark values chosen from the allowed parameter space of Fig. \ref{constraint-flavor}.}\label{benchmark}
\end{table}
\section{Inspection of new physics in $\Lambda_b \to \Lambda^* (1520)(\to pK^-)\ell^+\ell^-$ decays}
Having established the benchmark values for the new vector coefficient, we proceed to examine their effects on the decay observables of the exclusive $\Lambda_b \to \Lambda^* (1520)(\to pK^-)\ell^+\ell^-$ channel where $\ell = \mu, e$. This analysis aims to elucidate the impact of new physics parameters associated with the coupling $C_9^{NP}$ on the decay dynamics, thereby providing a deeper understanding of how new physics might influence this process. 
\subsection{Decay observables of $\Lambda_b \to \Lambda ^* (1520)(\to pK^-)\ell^+\ell^-$ process}
The four-fold angular distribution for the exclusive $\Lambda_b \to \Lambda ^*( \to pK^-)\ell^+\ell^-$ decay mode can be expressed as~\cite{Descotes-Genon:2019dbw, Das:2020cpv}
\begin{align}
\label{eq:fourfold}
\frac{d^4\mathcal{B}}{dq^2 d\cos \theta _ \ell d\cos \theta _{\Lambda ^*} d \phi} &= \frac{3}{8} \bigg[\bigg(\mathcal{K}_{1c}\cos \theta _ \ell + \mathcal{K}_{1cc}\cos ^2 \theta _ \ell + \mathcal{K}_{1ss}\sin ^2 \theta _\ell \bigg) \cos ^2 \theta _{\Lambda ^*}\, \nn\\ 
&~~~~+\bigg(\mathcal{K}_{2c}\cos \theta _ \ell + \mathcal{K}_{2cc}\cos ^2 \theta _ \ell + \mathcal{K}_{2ss}\sin ^2 \theta _\ell  \bigg)\sin^2 \theta_{\Lambda ^*}\, \nn\\
&~~~~+\bigg(\mathcal{K}_{3ss}\sin ^2 \theta _\ell \bigg)\sin ^2 \theta _{\Lambda ^*} \cos \phi + \bigg(\mathcal{K}_{4ss}\sin ^2 \theta _\ell \bigg)\sin ^2 \theta _{\Lambda ^*}\sin\phi\cos\phi \,\nn\\
&~~~~+\bigg(\mathcal{K}_{5s}\sin \theta _\ell  + \mathcal{K}_{5sc}\sin \theta _\ell \cos \theta _ \ell \bigg)\sin \theta _{\Lambda ^*}\cos \theta _{\Lambda ^*}\cos\phi\, \nn\\
&~~~~+\bigg(\mathcal{K}_{6s}\sin \theta _\ell  + \mathcal{K}_{6sc}\sin \theta _\ell \cos \theta _ \ell \bigg)\sin \theta _{\Lambda ^*}\cos \theta _{\Lambda ^*}\sin\phi
\bigg] ,
\end{align}
where $\theta_{\Lambda^{*}}$ represents the angle formed by the proton with the daughter baryon $\Lambda^{*}$ in the rest frame of
$\Lambda_b$. Similarly, in the rest frame of the lepton pair, $\theta_\ell$ denotes the angle formed by the $\ell^-$ with respect to the
direction of the daughter baryon $\Lambda^{*}$. Moreover, in the rest frame of $\Lambda_b$, $\phi$ defines the angle between the planes 
containing $p\,K^-$ and the lepton pair. 
The angular coefficients $\mathcal{K}_{1c, \cdots, 6sc}$, can be expressed as
\begin{equation}
\mathcal{K}_{1c, \cdots, 6sc} = {K}_{1c, \cdots, 6sc} + \frac{m_\ell}{\sqrt{q^2}} {K}^\prime_{1c, \cdots, 6sc} + \frac{m_\ell^2}{q^2} 
{K}^{\prime\prime}_{1c, \cdots, 6sc}\,.
\end{equation}
Here the first term ${K}$ corresponds to massless leptons, whereas, ${K}^{\prime}$ and ${K}^{\prime\prime}$ correspond to 
linear~($\mathcal{O}(m_\ell/\sqrt{q^2})$) and 
quadratic~($\mathcal{O}(m_\ell^2/q^2)$) mass 
corrections, respectively.
The explicit expressions for $K_{\{\cdots \}}$, ${K}^\prime_{\{\cdots \}}$ and ${K}^{\prime\prime}_{\{\cdots \}}$ in terms of 
transverse amplitude (Appendix \ref{AngC}) are taken Ref~\cite{Das:2020cpv}. 

From the differential decay distributions, one can construct several physical observables. 
 The differential branching ratio, the lepton forward-backward asymmetry, 
the fraction of longitudinal polarization and the ratio of branching fraction. These are given as follows 
\begin{itemize}
    \item The differential branching ratio \cite{Das:2020cpv}:
\end{itemize}
\begin{eqnarray}
\frac{d\mathrm{BR}}{dq^2} &=& \frac{1}{3}\bigg[K_{1cc} + 2K_{1ss} + 2K_{2cc} + 4K_{2ss} + 2K_{3ss} \bigg]
\end{eqnarray}
\begin{itemize}
    \item The lepton polarization asymmetry  \cite{Das:2020cpv}:
\end{itemize}
\begin{eqnarray}
    F_L(q^2) = 1 - \frac{2(K_{1cc} + 2K_{2cc} )}{K_{1cc} + 2(K_{1ss} + K_{2cc} + 2K_{2ss} + K_{3ss})}
\end{eqnarray}
\begin{itemize}
    \item The forward-backward asymmetry  \cite{Das:2020cpv}:
\end{itemize}
\begin{eqnarray}
    A^\ell_{\rm FB}(q^2) = \frac{3(K_{1c} + 2K_{2c} )}{2\big[ K_{1cc} + 2(K_{1ss} + K_{2cc} + 2K_{2ss} + K_{3ss} ) \big]}
\end{eqnarray}
\begin{figure}[thb]
\centering
\includegraphics[width=7.5cm,height=6cm]{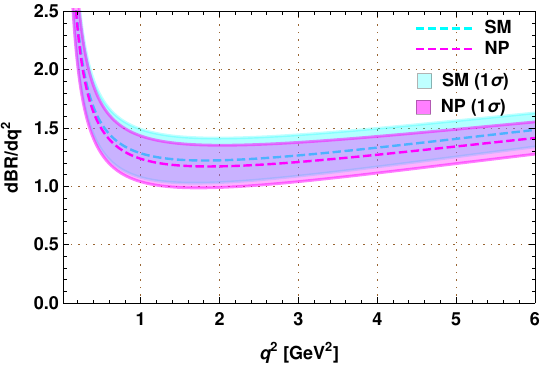}
\vspace{0.5 cm}
\includegraphics[width=7.5cm,height=6cm]{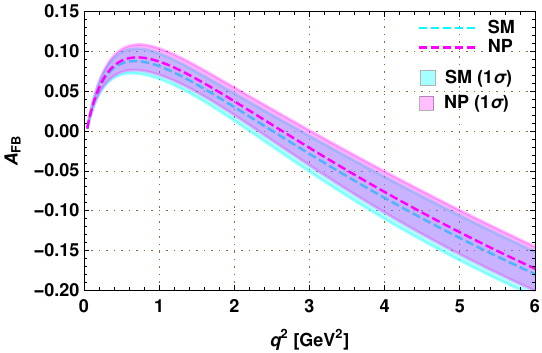}
\vspace{0.5 cm}
\includegraphics[width=7.5cm,height=6cm]{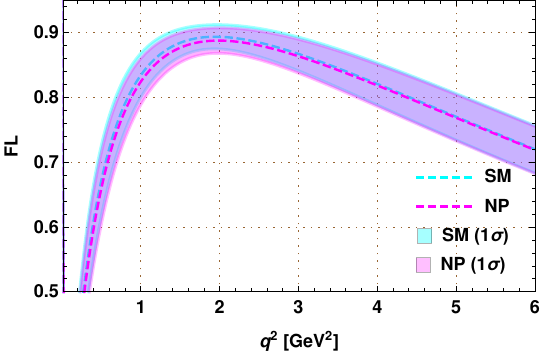}
\caption{Variation of the differential   branching ratio (in units of $10^{-9}$) (top left),  forward backward  asymmetry (top right), and Polarisation asymmetry (bottom)  of $\Lambda _b \to \Lambda ^* (1520) \mu ^+ \mu ^-$ process with $q^2$.}
\label{Observable plots}
\end{figure}
\begin{figure}[thb]
\centering
\includegraphics[width=8.15cm,height=6.85cm]{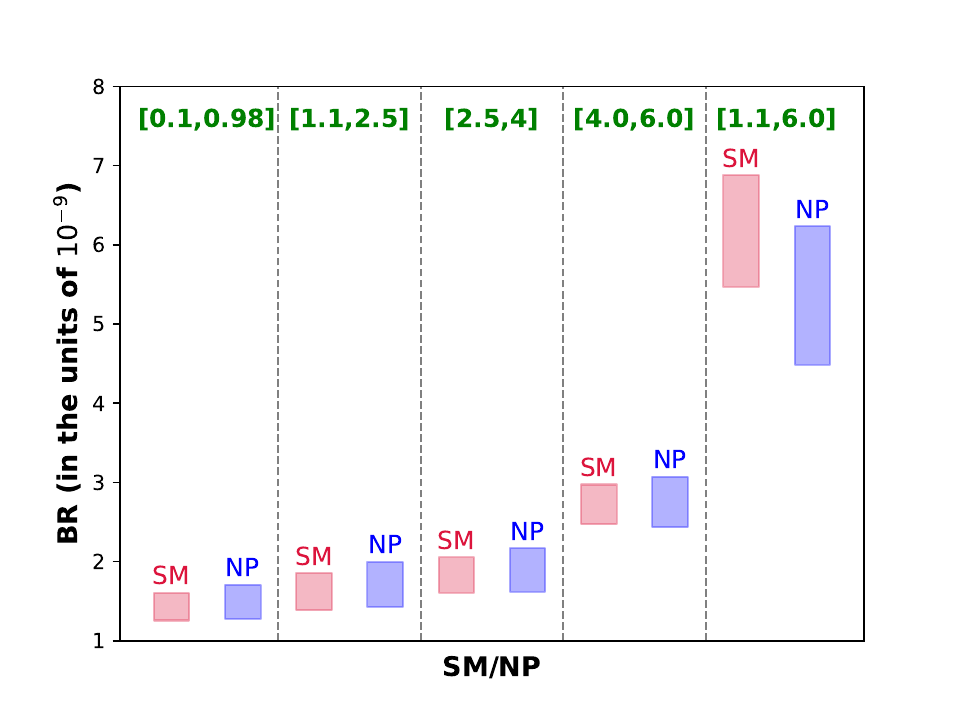}
\vspace{0.1 cm}
\includegraphics[width=8.15cm,height=6.85cm]{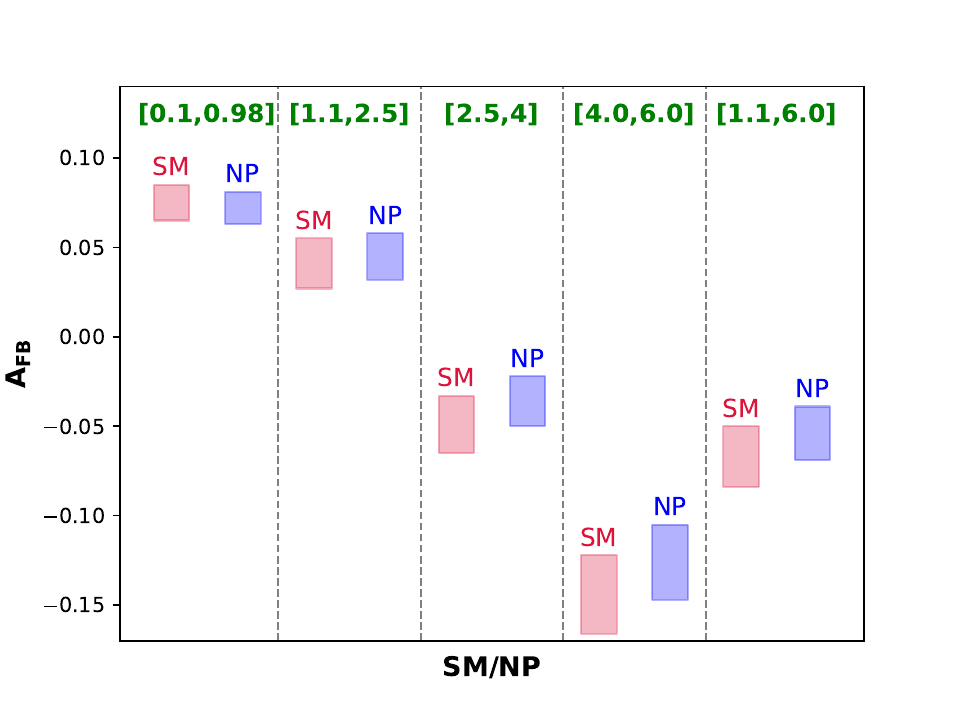}
\vspace{0.1 cm}
\includegraphics[width=8.75cm,height=6.85cm]{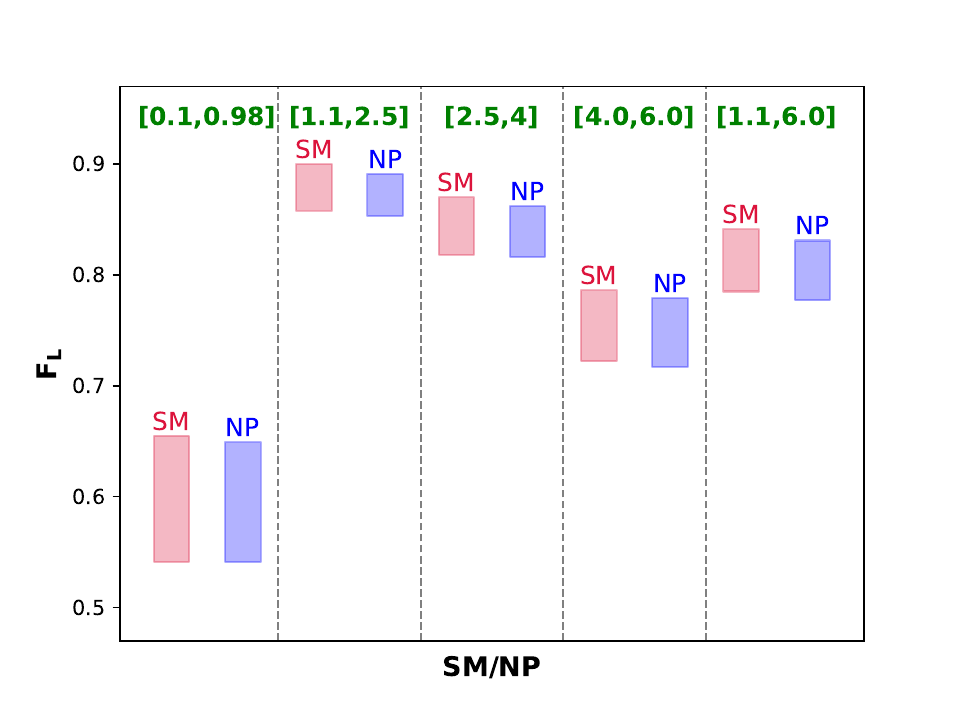}
\caption{ BR (top left),  $A_{FB}$ (top right) and $F_L$ (bottom panel)  of $\Lambda _b \to \Lambda ^* (1520) \mu ^+ \mu ^-$ process are shown across $q^2$ bins: $q^2 \in[0.1,0.98]$, $q^2 \in[1.1,2.5]$, $q^2 \in[2.5,4.0]$, $q^2 \in[4.0,6.0]$, $q^2 \in[1.1,6.0]$. The bands indicate $1\sigma$ uncertainty.}
\label{binwise plots}
\end{figure}

Using the input parameters from the Table~\ref{tab_input} and the light-front quark model form factor for the $\Lambda_b \to \Lambda^*$ transition as described in~\cite{Li:2022nim}, we investigate the observables associated with the $\Lambda_b \to \Lambda^*(1520) (\to pK^-)\ell^+\ell^-$ decay channel. We provide a detailed analysis and interpretation of our results below.
\subsection{Analysis of Results and Interpretations}
In this subsection, we conduct an in-depth analysis of the $\Lambda_b \to \Lambda^* (1520)(\to pK^-)\ell^+\ell^-$ decay process in the presence of new physics. With all the input parameters that are pertinent to our analysis, we display the variations of all the observables w.r.t $q^2$ in Fig. \ref{Observable plots}. Similarly, we show the corresponding $q^2$ bin-wise plots where we choose different bin sizes such as [0.1, 0.98], [1.1, 2.5], [2.5, 4] and [4.0, 6.0] (in the units of GeV$^2$) compatible with LHCb experiment. The color description of the $q^2$ dependency and bin-wise plots are as follows.
\begin{table}[htp]
\setlength{\tabcolsep}{10pt}
\centering
\caption{Numerical predictions for branching ratio and other physical  observables of $\Lambda_b \to \Lambda^* (1520)(\to pK^-)\ell^+\ell^-$  decay in SM and NP.}
\label{Tab:predictions}       
\begin{tabular}{|l| l l l|}
\hline
\multicolumn{4}{|c|}{\textbf{SM}} \\
\hline
$q^2$ bin & $<BR>_\mu\times10^{9}$ & $<A_{FB}>_\mu$ &$<F_L>_\mu$ 
\\\hline
$[0.1, 0.98]$ & $1.431 \pm 0.175 $  & $0.075 \pm 0.010$&$0.598 \pm 0.057$\\ 
$[1.1, 2.5]$&$1.621 \pm 0.236$  &$0.041 \pm 0.014$&$0.879 \pm 0.021$ \\
$[2.5, 4.0]$&$1.831 \pm 0.223$ &$-0.049 \pm 0.016$&$0.844 \pm 0.026$ \\
$[4.0, 6.0]$&$2.722 \pm 0.252$ &$-0.144 \pm 0.022$&$0.754 \pm 0.032$ \\
$[1.1, 6.0]$&$6.176 \pm 0.704$  &$-0.067 \pm 0.017$&$0.813 \pm  0.028$ \\
\hline
\multicolumn{4}{|c|}{\textbf{NP}} \\
\hline
$q^2$ bin  & $<BR>_\mu\times10^{9}$ & $<A_{FB}>_\mu$ & $<F_L>_\mu$
\\\hline
$[0.1, 0.98]$&$1.493 \pm 0.217$  &$0.072 \pm 0.009$&$0.595 \pm 0.054$ \\
$[1.1, 2.5]$&$1.712 \pm 0.284$ &$0.045 \pm 0.013$&$0.872 \pm 0.019$\\
$[2.5, 4.0]$&$1.895 \pm 0.274$  &$-0.036 \pm 0.014$&$0.839 \pm 0.023$ \\
$[4.0, 6.0]$&$2.753 \pm 0.320$ &$-0.126 \pm 0.021$&$0.748 \pm 0.031$ \\
$[1.1, 6.0]$&$6.362 \pm 0.875$  &$-0.054 \pm 0.015$&$0.804 \pm 0.027$ \\
\hline
\end{tabular}
\end{table}

\ding{43} Distribution Plot: The cyan dotted line represents the SM contribution, with the accompanying cyan band indicating the \(1\sigma\) error band arising from uncertainties in the form factors and CKM matrix elements. The magenta dotted line corresponds to the NP contribution, while the magenta band illustrates the \(1\sigma\) uncertainty associated with this contribution.

\ding{43} Bin-wise Plot: The SM and NP contributions are represented with crimson red and blue colors, respectively.

Now, the detailed observations in the presence of NP contribution are  provided as below.
\subsubsection{Branching ratio}
The top-left panel of Fig. \ref{Observable plots}, illustrates the $q^2$ dependence of the branching ratio for the $\Lambda_b \to \Lambda^* (1520)(\to pK^-)\mu^+\mu^-$ decay, both in the SM as well as including new physics contributions. In the presence of NP couplings, the branching ratio is found to be of the order $\mathcal{O}(10^{-9})$. The inclusion of the new vector coupling results in a reduction of BR($\Lambda_b \to \Lambda^* (1520)(\to pK^-)\mu^+\mu^-$), signifying a notable deviation from the SM prediction. The bin-wise plots are presented for both the  SM and with new physics in Fig. \ref{binwise plots}.
\subsubsection{Longitudinal polarization asymmetry}
Due to the influence of the  new physics the longitudinal polarization asymmetry ($F_L$) is shifted slightly lower compared to the SM. However, no significant deviation is observed for this particular observable. The corresponding $q^2$-distribution and bin-wise plots are shown in the bottom-left panel of Fig. \ref{Observable plots} and Fig. \ref{binwise plots}, respectively. In Table \ref{Tab:predictions}, we provide detailed numerical predictions for both the SM and new physics, offering a comprehensive comparison of theoretical expectations. 
\subsubsection{Forward-backward asymmetry}
This observable exhibits a comparatively larger deviation compared to branching ratio and polarization asymmetry. The $q^2$ distribution, illustrated in Fig. \ref{Observable plots}, reveals a zero-crossing point at approximately 2.5 GeV$^2$ in the SM. However, in the presence of new physics couplings, the zero-crossing point shifts to around 2.8 GeV$^2$, a deviation that is distinctly noticeable compared to the Standard Model value at around 2.5 GeV$^2$.

From our investigation, the SM and NP predictions more or less consistent with each other. 
Nevertheless, ongoing and future efforts are expected to improve sensitivity to rare baryonic decays could disentangle the  SM and NP predictions. Advances in reconstruction techniques, increased luminosity, and refined selection strategies may enable the precise measurement of these observables with the required precision. While this remains speculative at this stage, our analysis serves as a theoretical motivation for pursuing such measurements, as a precise determination of these observables could provide valuable insights into possible NP contributions.
\section{Concluding Remarks}
Motivated by the observed anomalies in $B \to (K, K^*) \mu^+ \mu^-$ and $B_s \to \phi \mu^+ \mu^-$ decays, which proceed via the $b \to s \mu^+ \mu^-$ flavor-changing neutral current interaction, we examine the exclusive semileptonic $\Lambda_b \to \Lambda^* (1520)(\to pK^-)\mu^+\mu^-$ decay channel in the context of a $U(1)_{L_e-L_\mu}$ gauge extension. This extended model incorporates an enriched particle content, including three neutral fermions, the lightest of which contributes to the dark matter content of the Universe. Additionally, the scalar sector is augmented by a $\tilde{R}_2$ scalar leptoquark doublet to investigate flavor decays in the $B$-meson sector. Focusing on the muon mode in the final lepton pair, we constrain the NP coupling by considering the global analysis associated with observables of $B \to K^{(*)} \mu^+ \mu^-$ and $B_s \to \phi \mu^+ \mu^-$ decay channels. Furthermore the fermionic dark matter annihilates via scalar leptoquark components and $Z^\prime$ to provide the relic abundance of dark matter. DM-nucleon interaction gives spin-dependent cross section in leptoquark portal and spin-independent interaction through Higgs portal. Dark matter observables constrain the  model parameters and the constrained parameter space is utilized in the flavor studies. Utilizing the allowed parameter space consistent with both flavor and dark matter sectors, we explore the impact on various observables, including the branching ratio, forward-backward asymmetry and polarization asymmetry in the $\Lambda_b \to \Lambda^* (1520) (\to pK) \ell^+ \ell ^-$ decay channel.

Our analysis reveals that the differential branching ratio is reduced compared to the SM, with a minimal deviation. We observe a substantial contribution from new physics in the analysis of the forward-backward asymmetry. 
It is crucial to gather more data from these experiments to fully understand the significance of the new physics contributions.
\acknowledgments 
MKM acknowledges the financial support from IoE PDRF, University of Hyderabad. SS acknowledges the support of IoE project, University of Hyderabad and also IACS Kolkata istitute funding for Research Associate-I. SS would like to thank Prof. Sourov Roy for the hospitality provided in their research lab. DP extends appreciation for the support of Prime Minister's Research Fellowship, Government of India. RM would like to thank University of Hyderabad IoE project grant no. RC1-20-012. MKM extends sincere thanks to Dr. Suchismita Sahoo for the essential assistance in this work, also to Dr. Jacky Kumar and Dr. Girish Kumar for their valuable suggestions concerning the \textit{flavio} package. We would like to thank  Papia Panda and Priya Mishra for useful discussions.
\appendix  
\section{Angular Coefficients} \label{AngC}
The expression of the angular coefficients is
\begin{align}
    \mathcal{K}_{1c} &= -2\beta_\ell \bigg(\re(A_{\perp1}^L A_{\parallel1}^{L*}) - \left\{ \text{L} \leftrightarrow \text{R} \right\} \bigg)\, ,\\
    \mathcal{K}_{1c}^\prime&=-2\beta_\ell\bigg(\re(\apasl\apaol^{\ast})+\re(\apasr\apaol^{\ast})+\re(\apesl\apeol^{\ast})+\re(\apesr\apeol^{\ast}) + \{L\leftrightarrow R\} \bigg)\,  ,\\
    \mathcal{K}_{1c}^{\prime\prime} &= 0\, ,\\
    \mathcal{K}_{1cc} &= \bigg(|A_{\parallel 1}^L|^2 + |A_{\parallel S}^L|^2 + |A_{\perp1}^L|^2 + |A_{\perp S}^L|^2 
    + \left\{ \text{L} \leftrightarrow \text{R} \right\} \bigg)\, ,\\
    \mathcal{K}_{1cc}^\prime &= 2\bigg(-\re(A_{\parallel t}^R A_{\parallel S}^{L*}) + \re(A_{\parallel S}^L A_{\parallel t}^{L*}) 
    - \re(A_{\perp t}^R A_{\perp S}^{L*}) + \re(A_{\perp S}^L A_{\perp t}^{L*}) + \left\{ \text{L} \leftrightarrow \text{R} \right\} \bigg)\, ,\\
    \mathcal{K}_{1cc}^{\prime\prime} &= 2 \bigg(|\apaol|^2 -|\apanl|^2-|\apasl|^2+|\apatl|^2+|\apeol|^2-|\apenl|^2-|\apesl|^2+|\apetl|^2\nn\\\
&\quad\quad+\re(\apaor\apaol^{\ast})+\re(\apanr\apanl^{\ast})-\re(\apasr\apasl^{\ast})-\re(\apatr\apatl^{\ast})\nn\\\
&\quad\quad+\re(\apeor\apeol^{\ast})+\re(\apenr\apenl^{\ast})-\re(\apesr\apesl^{\ast})-\re(\apetr\apetl^{\ast})+ \{ L\leftrightarrow R\} \bigg)\, ,\\
    \mathcal{K}_{1ss} &= \frac{1}{2} \bigg(2 |A_{\parallel 0}^L|^2 + |A_{\parallel 1}^L|^2 + 2 |A_{\parallel S}^L|^2 
    + 2 |A_{\perp 0}^L|^2 + |A_{\perp 1}^L|^2 + 2 |A_{\perp S}^L|^2 + \left\{ L \leftrightarrow R \right\} \bigg)\, ,\\
    \mathcal{K}^{\prime}_{1ss} &= -2 \bigg( \re(A_{\parallel t}^R A_{\parallel S}^{L*}) - \re(A_{\parallel S}^L A_{\parallel t}^{L*}) 
    + \re(A_{\perp t}^R A_{\perp S}^{L*}) - \re(A_{\perp S}^L A_{\perp t}^{L*}) + \left\{ \text{L} \leftrightarrow \text{R} \right\} \bigg)\, ,\\
    \mathcal{K}_{1ss}^{\prime\prime} &= 2 \bigg(-|\apaol|^2 - |\apasl|^2 + |\apatl|^2 - |\apeol|^2 - |\apesl|^2 + |\apetl|^2 \nonumber \\
    &\quad + \re(\apaor\apaol^{\ast}) + \re(\apanr\apanl^{\ast}) - \re(\apasr\apasl^{\ast}) - \re(\apatr\apatl^{\ast}) \nonumber \\
    &\quad + \re(\apeor\apeol^{\ast}) + \re(\apenr\apenl^{\ast}) - \re(\apesr\apesl^{\ast}) - \re(\apetr\apetl^{\ast}) 
    + \left\{ L\leftrightarrow R \right\} \bigg)\, ,\\
\mathcal{K}_{2c} &= -\frac{1}{2} \beta_\ell \bigg( \re(\apenl\apanl^{\ast}) + 3\re(\bpenl\bpanl^{\ast}) - \{ L\leftrightarrow R \} \bigg)\, ,\nn\\
    \mathcal{K}_{2c}^{\prime} &= -\frac{1}{2} \beta_\ell \bigg( \re(\apasl\apaol^{\ast}) + \re(\apasl\apaor^{\ast}) + \re(\apesl\apeol^{\ast}) + \re(\apesl\apeor^{\ast}) + \{ L\leftrightarrow R \} \bigg)\, ,\nn\\
    \mathcal{K}_{2c}^{\prime\prime} &= 0\, ,\\
    \mathcal{K}_{2cc} &= \frac{1}{4} \bigg( |\apanl|^2 + |\apasl|^2 + 3|\bpanl|^2 + |\apenl|^2 + |\apesl|^2 + 3|\bpenl|^2 + \{ L\leftrightarrow R \} \bigg)\, ,\nn\\
    \mathcal{K}_{2cc}^\prime &= -\frac{1}{2} \bigg( \re(\apatr\apasl^{\ast}) - \re(\apasl\apatl^{\ast}) + \re(\apetr\apesl^{\ast}) - \re(\apesl\apetl^{\ast})+ \{ L\leftrightarrow R \} \bigg)\, ,\nn\\
    \mathcal{K}_{2cc}^{\prime\prime} &= \frac{1}{2} \bigg( |\apaol|^2 - |\apanl|^2 - |\apasl|^2 + |\apatl|^2 + |\apeol|^2 - |\apenl|^2 - |\apesl|^2 + |\apetl|^2 \nn\\
    &\quad\quad -3|\bpanl|^2 -3|\bpenl|^2 + \re(\apaor\apaol^{\ast}) + \re(\apanr\apanl^{\ast}) - \re(\apasr\apasl^{\ast}) - \re(\apatr\apatl^{\ast})\nn\\
    &\quad\quad + \re(\apeor\apeol^{\ast}) + \re(\apenr\apenl^{\ast}) - \re(\apesr\apesl^{\ast}) - \re(\apetr\apetl^{\ast})\nn\\
    &\quad\quad + 3\re(\bpanr\bpanl^{\ast}) + 3\re(\bpenr\bpenl^{\ast}) + \{L\leftrightarrow R\} \bigg)\, ,\\
    \mathcal{K}_{2ss} &= \frac{1}{8}\bigg(2|\apaol|^2+|\apanl|^2+2|\apasl|^2+2|\apeol|^2+|\apenl|^2+2|\apesl|^2 \nonumber \\
&\quad\quad+3|\bpanl|^2+3|\bpenl|^2-2\sqrt{3}\re(\bpanl\apanl^{\ast})+2\sqrt{3}\re(\bpenl\apenl^{\ast}) + \{L\leftrightarrow R\}\bigg)\, ,\nonumber \\ 
K_{2ss}^\prime &= -\frac{1}{2}\bigg(\re(\apatr\apasl^{\ast})-\re(\apasl\apatl^{\ast})+\re(\apetr\apesl^{\ast})-\re(\apesl\apetl^{\ast})+\{L\leftrightarrow R\}\bigg)\, ,\nonumber \\ 
\mathcal{K}_{2ss}^{\prime\prime} &= \frac{1}{2}\bigg(-|\apaol|^2-|\apasl|^2+|\apatl|^2-|\apeol|^2-|\apesl|^2+|\apetl|^2 \nonumber \\
&\quad\quad+\re(\apaor\apaol^{\ast})+\re(\apanr\apanl^{\ast})+2\sqrt{3} \re(\bpanl\apanl^{\ast})-\re(\apasr\apasl^{\ast}) \nonumber \\
&\quad\quad-\re(\apatr\apatl^{\ast})+\re(\apeor\apeol^{\ast})+\re(\apenr\apenl^{\ast})-2\sqrt{3}\re(\bpenl\apenl^{\ast})-\re(\apesr\apesl^{\ast}) \nonumber \\
&\quad\quad-\re(\apetr\apetl^{\ast})
+ 3\re(\bpanr\bpanl^{\ast})+3\re(\bpenr\bpenl^{\ast})+ \{L\leftrightarrow R \}\bigg)\, ,\\ 
\mathcal{K}_{3ss} &= \frac{\sqrt{3}}{2}\bigg(\re(\bpanl\apanl^{\ast})-\re(\bpenl\apenl^{\ast})+\{L\leftrightarrow R\}\bigg)\, ,\nonumber \\ 
\mathcal{K}_{3ss}^{\prime} &= 0\, ,\nonumber \\ 
\mathcal{K}_{3ss}^{\prime\prime} &= -2\sqrt{3}\bigg(\re(\bpanl\apanl^{\ast})-\re(\bpenl\apenl^{\ast})+\{L\leftrightarrow R\}\bigg)\, ,\\ 
\mathcal{K}_{4ss} &= \frac{\sqrt{3}}{2}\bigg(\im(\bpenl\apanl^{\ast})-\im(\bpanl\apenl^{\ast})+\{L\leftrightarrow R\}\bigg)\, ,\nonumber \\ 
\mathcal{K}_{4ss}^{\prime} &= 0\, ,\nonumber \\ 
\mathcal{K}_{4ss}^{\prime\prime} &= -2\sqrt{3}\bigg(\im(\bpenl\apanl^{\ast})-\im(\bpanl\apenl^{\ast})+\{L\leftrightarrow R\}\bigg)\, ,\\ 
\mathcal{K}_{5s} &= \frac{\sqrt{6}}{2}\beta_\ell\bigg(\re(\bpenl\apaol^{\ast})-\re(\bpanl\apeol^{\ast})-\{L \leftrightarrow R\}\bigg)\, ,\nonumber \\ 
\mathcal{K}_{5s}^{\prime} &= -\frac{\sqrt{6}}{2}\beta_\ell\bigg(\re(\bpanr\apasl^{\ast})-\re(\bpenr\apesl^{\ast})
+\re(\apasl\bpanl^{\ast})-\re(\apesl\bpenl^{\ast})
+\{ L\leftrightarrow R\}\bigg)\, ,\nonumber \\ 
\mathcal{K}_{5s}^{\prime\prime} &= 0\, ,\\ 
\mathcal{K}_{5sc} &= -\frac{\sqrt{6}}{2}\bigg(\re(\bpanl\apaol^{\ast})-\re(\bpenl\apeol^{\ast})+\{L \leftrightarrow R\}\bigg)\, ,\nonumber \\ 
\mathcal{K}_{5sc}^{\prime} &= 0\, ,\\ 
\mathcal{K}_{5sc}^{\prime\prime} &= 2\sqrt{6}\bigg(\re(\bpanl\apaol^{\ast})-\re(\bpenl\apeol^{\ast})+\{L \leftrightarrow R\}\bigg)\, ,\nonumber \\ 
\mathcal{K}_{6s} &= \frac{\sqrt{6}}{2}\beta_\ell\bigg(\im(\bpanl\apaol^{\ast})-\im(\bpenl\apeol^{\ast})-\{ L\leftrightarrow R\}\bigg)\, ,\nonumber \\ 
\mathcal{K}_{6s}^{\prime} &= -\frac{\sqrt{6}}{2}\beta_\ell\bigg(\im(\bpenr\apasl^{\ast})-\im(\bpanr\apesl^{\ast})+\im(\apesl\bpanl^{\ast})-\im(\apasl\bpenl^{\ast})
+\{ L\leftrightarrow R\}\bigg)\, ,\\ 
\mathcal{K}_{6s}^{\prime\prime} &= 0\, ,\nonumber \\ 
\mathcal{K}_{6sc} &= -\frac{\sqrt{6}}{2}\bigg(\im(\bpenl\apaol^{\ast})-\im(\bpanl\apeol^{\ast})+\{L\leftrightarrow R\}\bigg)\, ,\nonumber \\ 
\mathcal{K}_{6sc}^{\prime} &= 0\, ,\nonumber \\ 
\mathcal{K}_{6sc}^{\prime\prime} &= 2\sqrt{6}\bigg(\im(\bpenl\apaol^{\ast})-\im(\bpanl\apeol^{\ast})+\{L \leftrightarrow R\}\bigg)\, .
\end{align}
\section{Loop Functions}  \label{loop}

The loop functions essential for calculating the $B_s - \bar{B}_s$ mixing take the following form:  

\begin{equation} \label{A:loop-2}  
f(\chi_1, \chi_2, \chi_3,\dots) \equiv \frac{f(\chi_1, \chi_3,\dots) - f(\chi_2, \chi_3,\dots)}{\chi_1 - \chi_2},~~~~~~~~ f = j, \kappa\,,  
\end{equation}  

with  

\begin{align} \label{A:loop-3}  
j(\chi) &= \frac{\chi \log \chi}{\chi - 1}\,,\\\label{A:loop-4}  
\kappa(\chi) &= \frac{\chi^2 \log \chi}{\chi - 1}\,.  
\end{align}  

\bibliographystyle{apsrev4-1}
\bibliography{BL}

\end{document}